%% file: main.tex
\documentclass[10pt,twocolumn,letterpaper]{article}

\usepackage{cvpr}              

\input{preamble}

\definecolor{cvprblue}{rgb}{0.21,0.49,0.74}
\usepackage[pagebackref,breaklinks,colorlinks,citecolor=cvprblue]{hyperref}

\def\paperID{46} 
\def\confName{3DV\xspace}
\def\confYear{2026\xspace}

\title{Inverse Rendering for High-Genus Surface Meshes from Multi-View Images}

\author{Xiang Gao\\
Futurewei Technologies\\
Stony Brook University\\
{\tt\small gao2@cs.stonybrook.edu}
\and
Xinmu Wang\\
Futurewei Technologies\\
Stony Brook University\\
{\tt\small xinmuwang@cs.stonybrook.edu}
\and
Xiaolong Wu\\
Futurewei Technologies\\
Purdue University\\
{\tt\small wu1565@purdue.edu}
\and
Jiazhi Li\\
Futurewei Technologies\\
University of Southern California\\
{\tt\small jiazhil@usc.edu}
\and
Jingyu Shi\\
Futurewei Technologies\\
Purdue University\\
{\tt\small shi537@purdue.edu}
\and
Yu Guo\\
Futurewei Technologies\\
George Mason University\\
{\tt\small tflsguoyu@gmail.com}
\and
Yuanpeng Liu\\
Futurewei Technologies\\
Stony Brook University\\
{\tt\small yuanpliu@cs.stonybrook.edu}
\and
Xiyun Song\\
Futurewei Technologies\\
{\tt\small xsong@futurewei.com}
\and
Heather Yu\\
Futurewei Technologies\\
{\tt\small hyu@futurewei.com}
\and
Zongfang Lin\\
Futurewei Technologies\\
{\tt\small zlin1@futurewei.com}
\and
Xianfeng David Gu\\
Stony Brook University\\
{\tt\small gu@cs.stonybrook.edu}
}

\input{sec/preamble}

\begin{document}
\twocolumn[{%
\renewcommand\twocolumn[1][]{#1}%
\maketitle

\input{fig/teaser}
}]

\input{sec/0_abstract}
\input{sec/1_intro}

\input{sec/2_related}

\input{sec/3_background}

\input{sec/4_method}

\input{sec/5_results}

\input{sec/6_conclusion}


{
    \small
    \bibliographystyle{ieeenat_fullname}
    \bibliography{main}
}

\end{document}

%% file: preamble.tex
\usepackage[dvipsnames]{xcolor}


%% file: sec/preamble.tex
\usepackage{amsmath}
\usepackage{amssymb}
\usepackage{graphicx}
\usepackage{overpic}
\usepackage{multirow}
\usepackage{pict2e}  
\usepackage{makecell}  
\usepackage[ruled,vlined]{algorithm2e}
\usepackage[table]{xcolor}

\usepackage[capitalize]{cleveref}
\crefname{section}{Sec.}{Secs.}
\Crefname{section}{Section}{Sections}
\Crefname{table}{Table}{Tables}
\crefname{table}{Tab.}{Tabs.}

\newlength{\resLen}

\newcommand\imwidth{}
\newcommand\imheight{}

\newcommand\xfou{}
\newcommand\yfou{}
\newcommand\wfou{}
\newcommand\hfou{}

\newcommand\leftfou{}
\newcommand\botfou{}
\newcommand\rightfou{}
\newcommand\topfou{}
\newcommand\bxfou{}
\newcommand\bxxfou{}
\newcommand\byfou{}
\newcommand\byyfou{}

%% file: fig/teaser.tex
\renewcommand\imwidth{4171}
\renewcommand\imheight{4171}
\renewcommand\xfou{1550}
\renewcommand\yfou{1550}
\renewcommand\wfou{250}
\renewcommand\hfou{250}

\renewcommand\leftfou{\fpeval{\xfou}}
\renewcommand\botfou{\fpeval{\imheight-\yfou-\hfou}}
\renewcommand\rightfou{\fpeval{\imwidth-\xfou-\wfou}}
\renewcommand\topfou{\fpeval{\yfou}}
\renewcommand\bxfou{\fpeval{\xfou/\imwidth*100}}
\renewcommand\bxxfou{\fpeval{(\xfou+\wfou)/\imwidth*100}}
\renewcommand\byfou{\fpeval{(\imheight-\yfou-\hfou)/\imwidth*100}}
\renewcommand\byyfou{\fpeval{(\imheight-\yfou)/\imwidth*100}}

\setlength{\resLen}{0.15\linewidth}
\addtolength{\tabcolsep}{-4pt}
\hskip -5pt
\begin{tabular}{cccccccc}
     & \raisebox{30pt}{\thead{Nicolet\\et al.~\cite{Nicolet2021Large}}} &
    \begin{overpic}[height=\resLen]{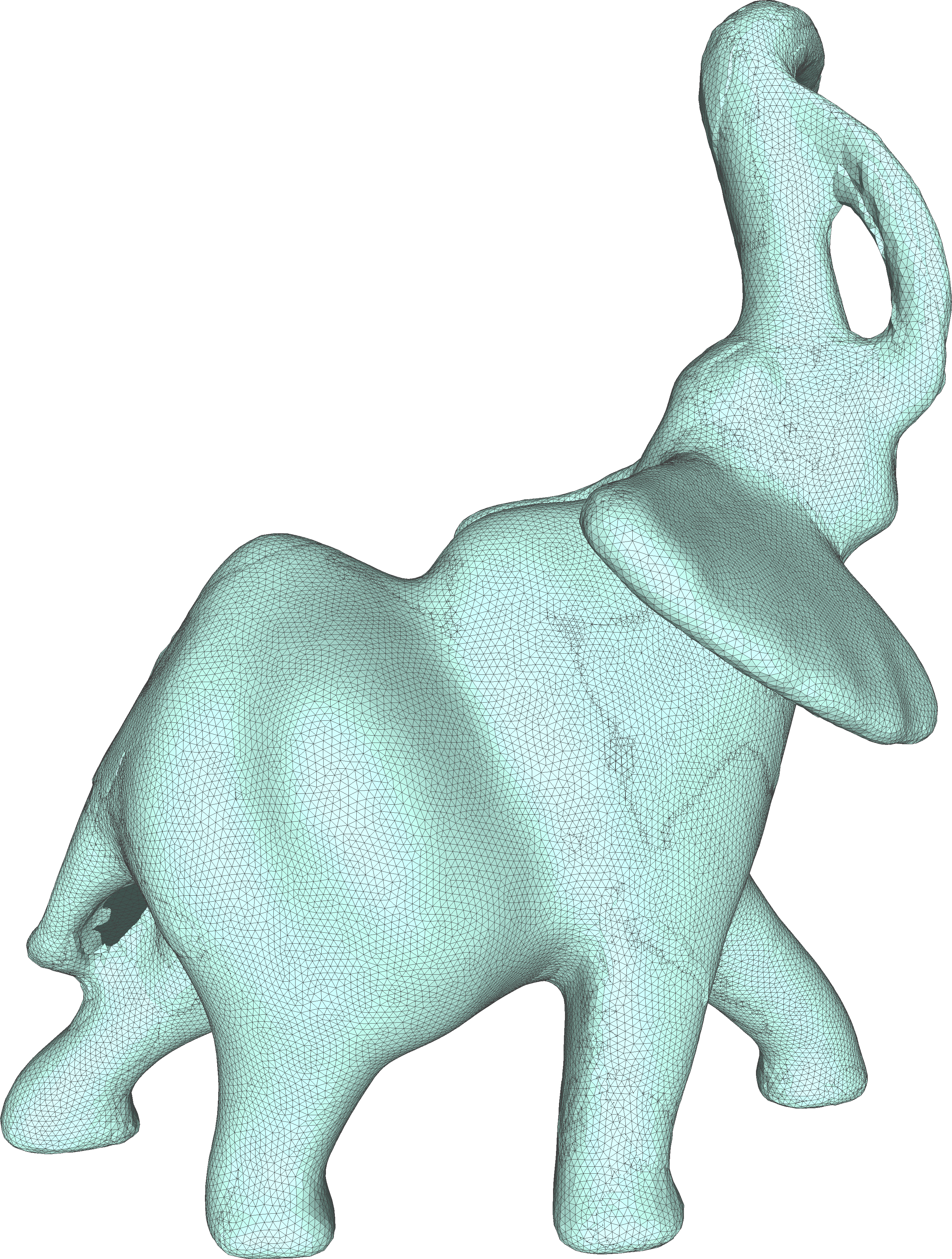}
    \end{overpic} &
    \begin{overpic}[height=\resLen]{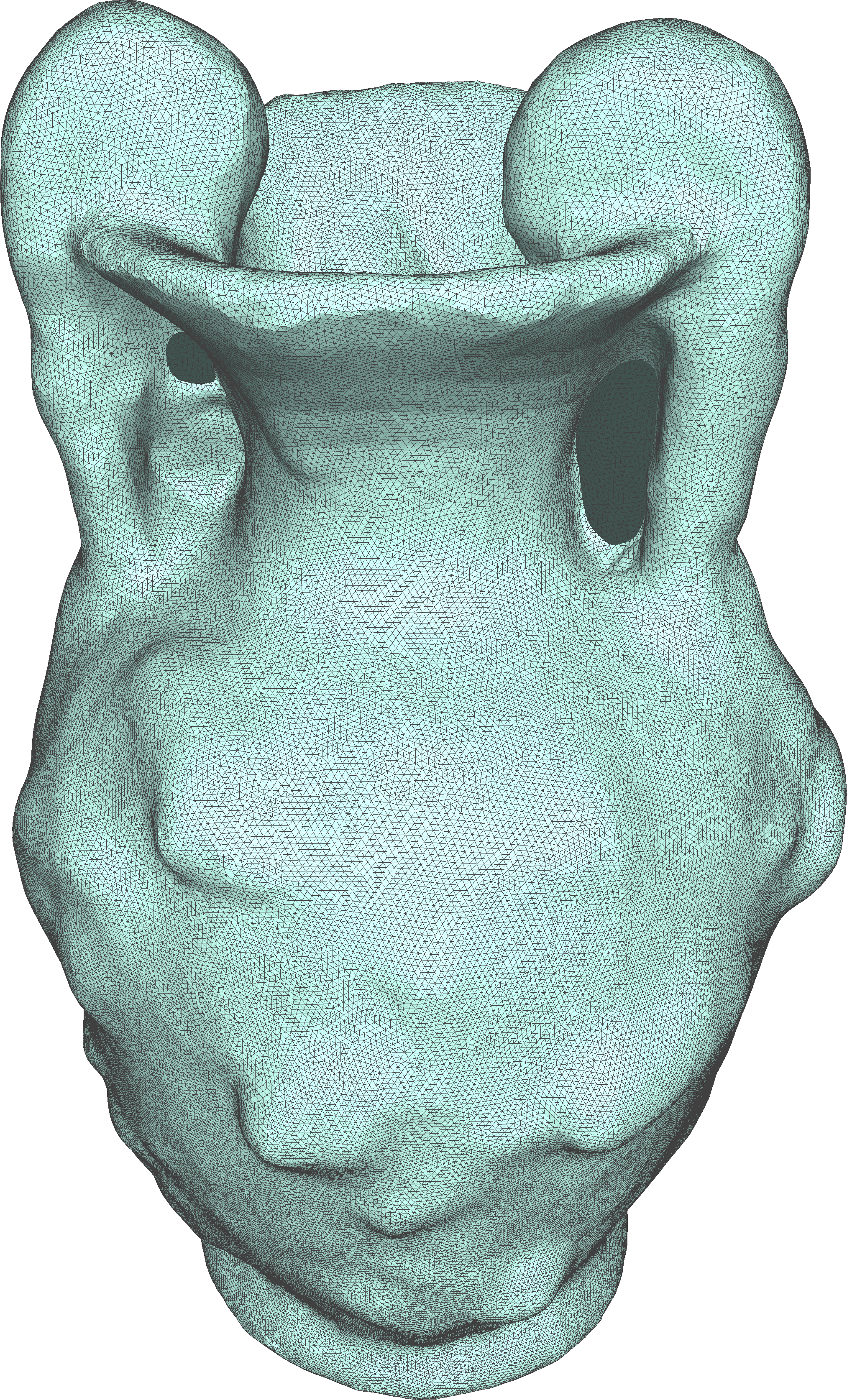}
    \end{overpic} &
    \begin{overpic}[height=\resLen]{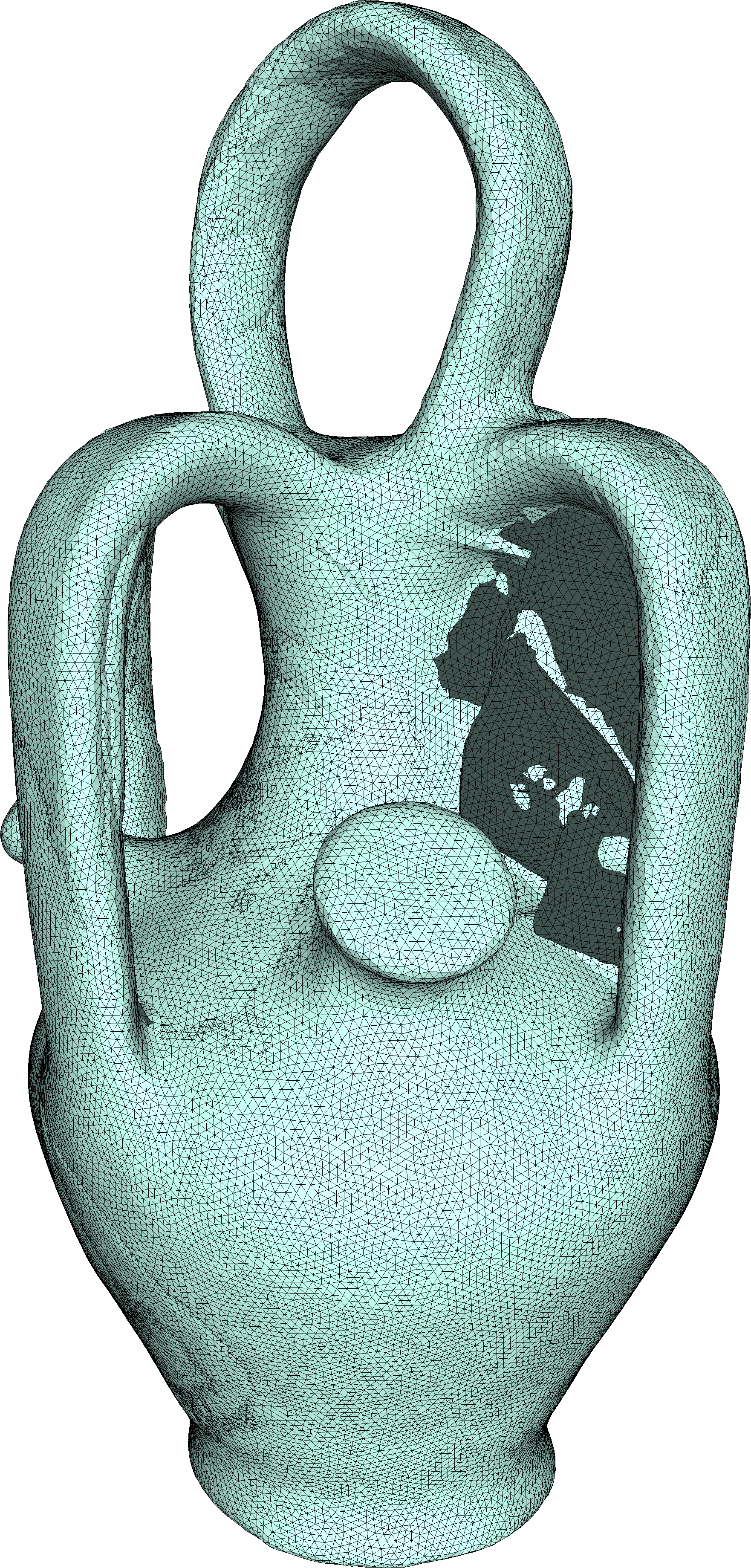}
    \end{overpic} &
    \begin{overpic}[height=\resLen]{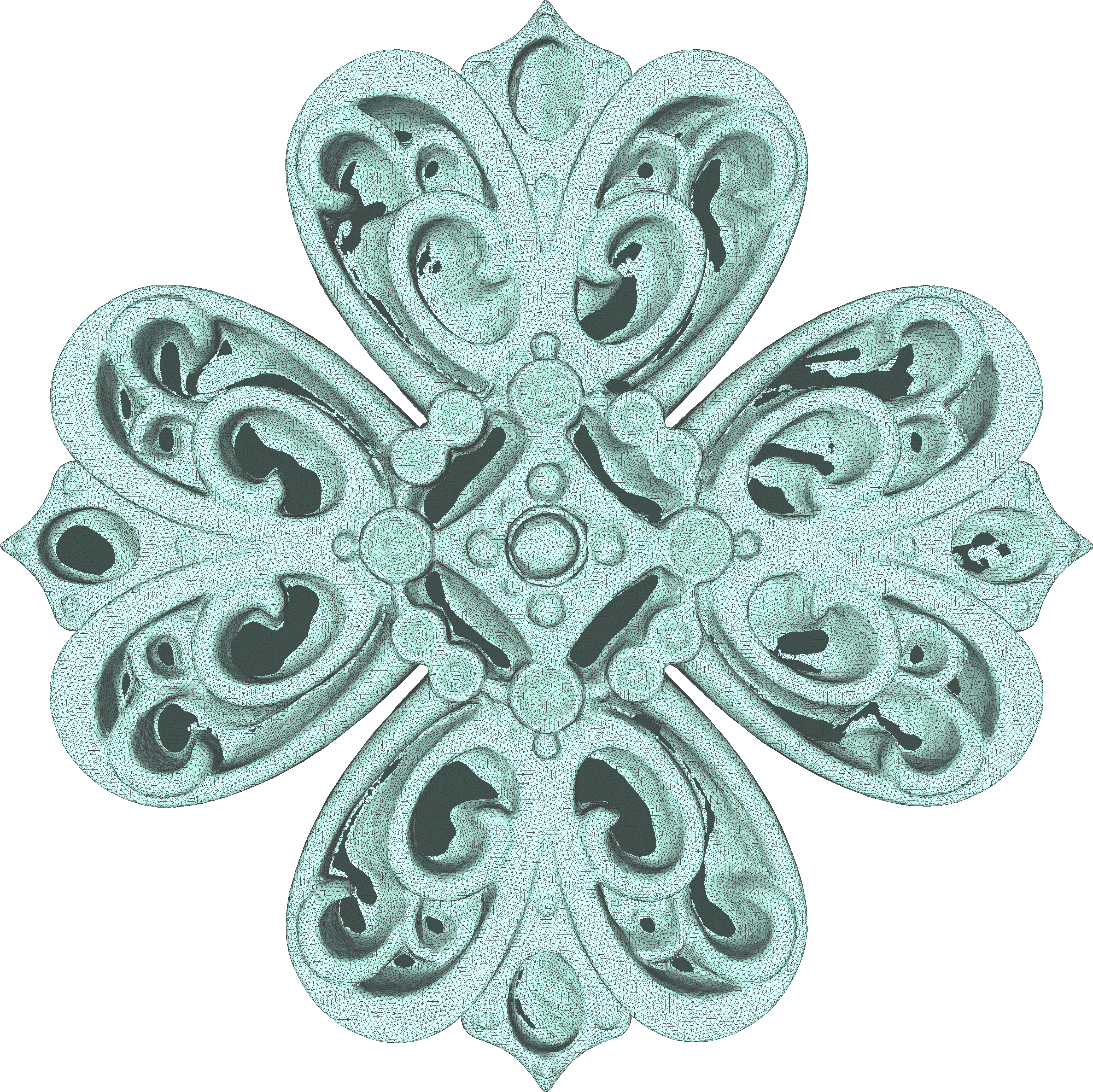}
    \end{overpic} &
    \begin{overpic}[height=\resLen]{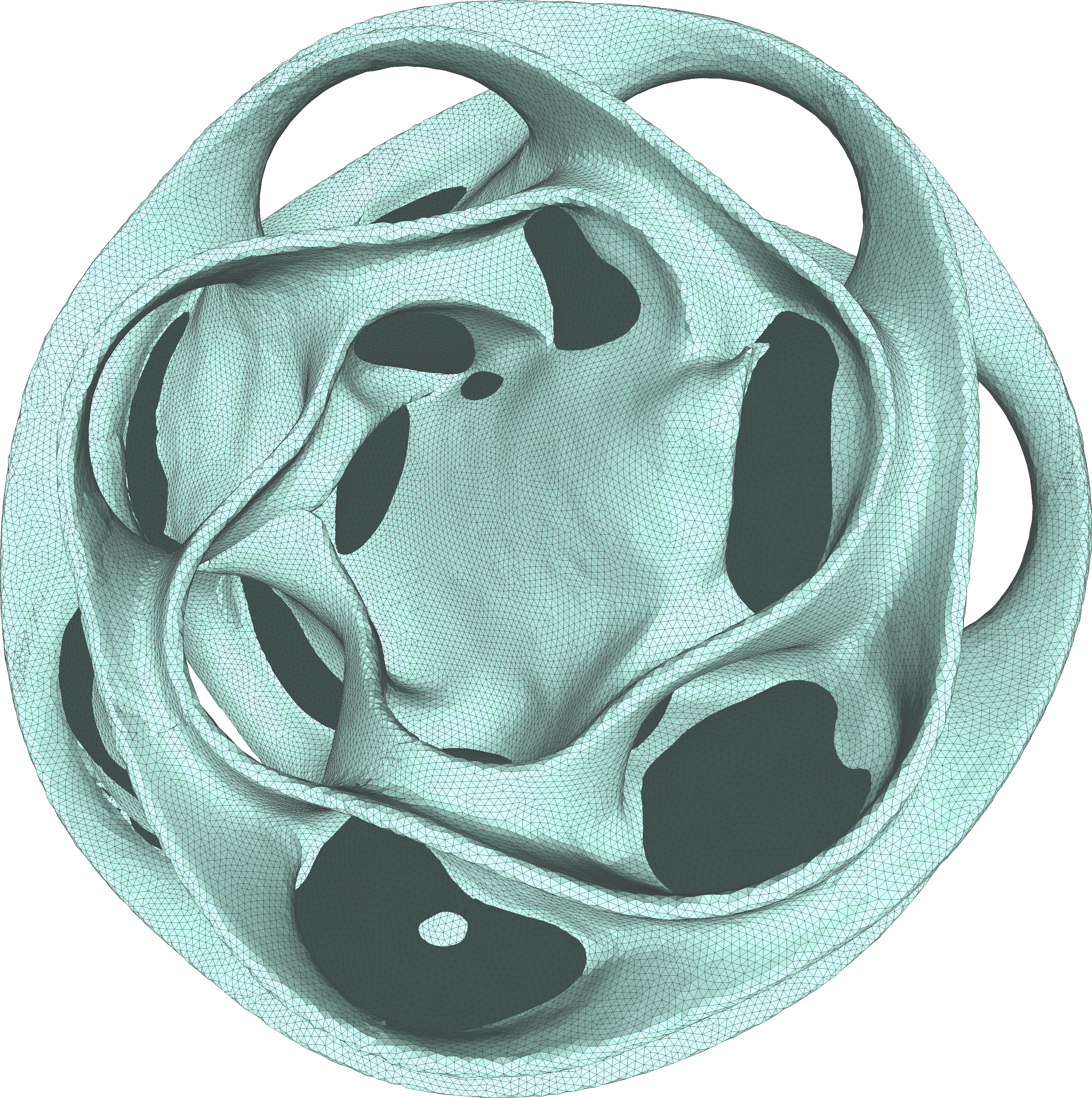}
    \end{overpic} &
    \begin{overpic}[height=\resLen]{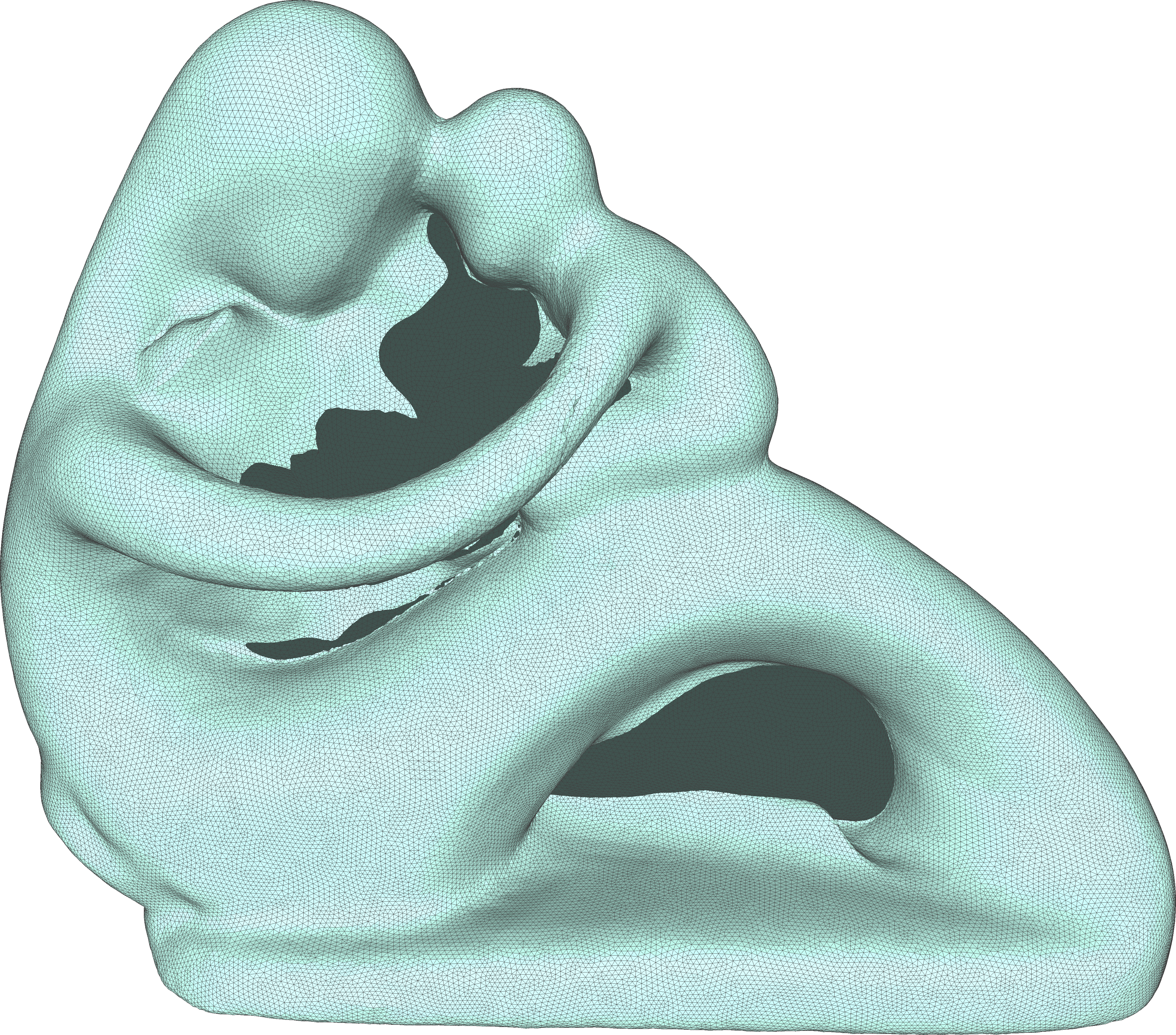}
    \end{overpic} 
    \\
    \raisebox{20pt}{
        \rule{0.06\linewidth}{0pt} 
    }
     & \raisebox{30pt}{Ours} &
    \begin{overpic}[height=\resLen]{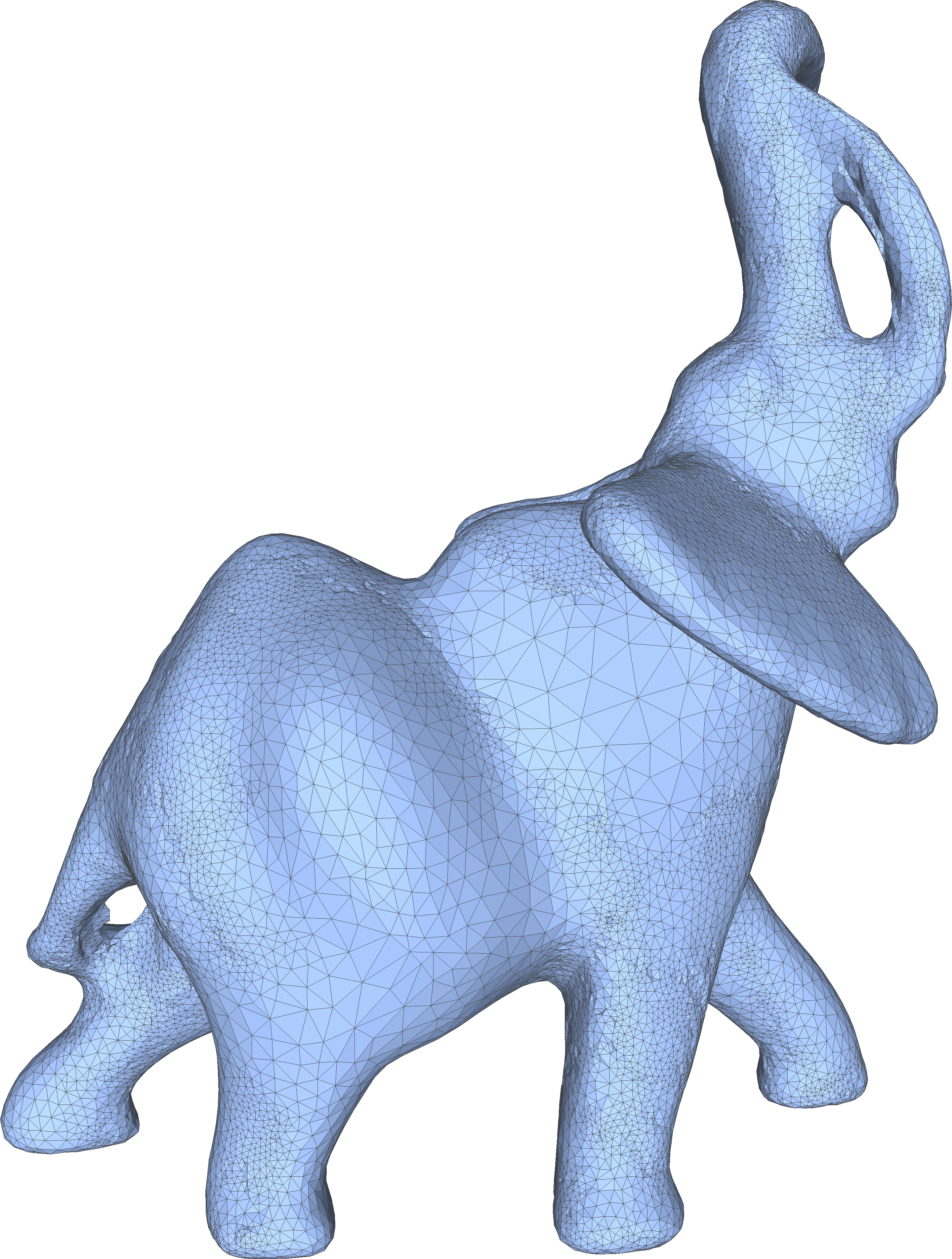}
    \end{overpic} &
    \begin{overpic}[height=\resLen]{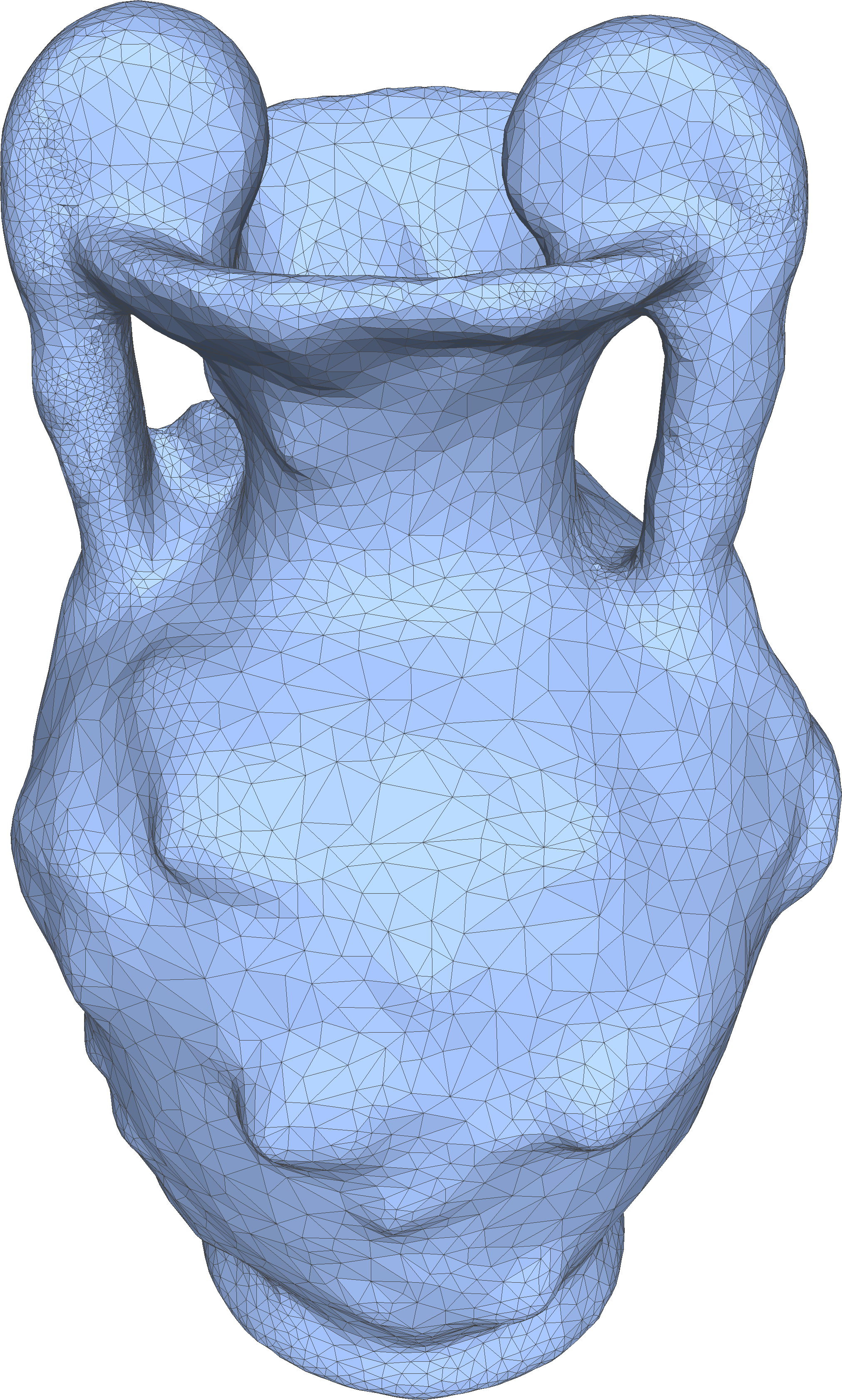}
    \end{overpic} &
    \begin{overpic}[height=\resLen]{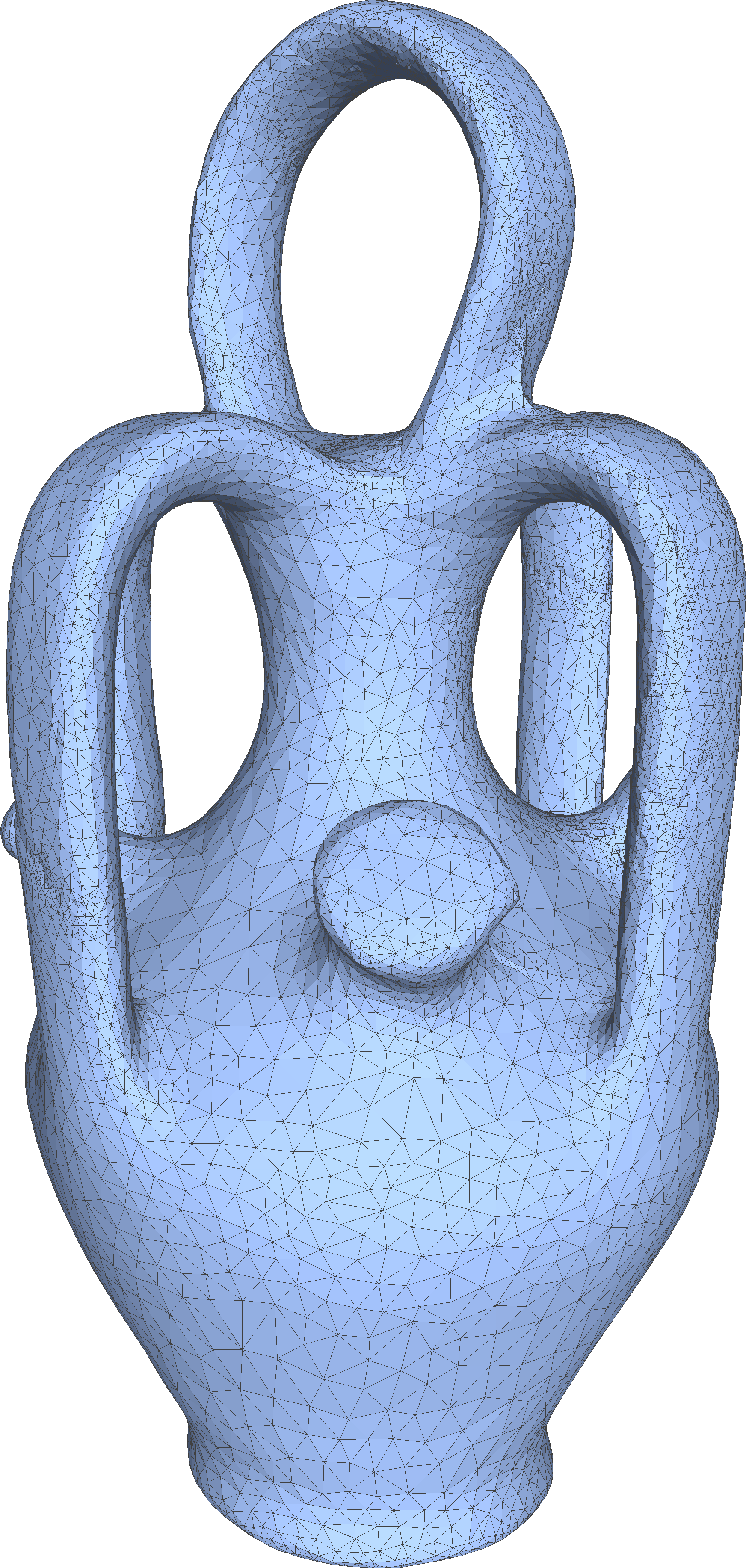}
    \end{overpic} &
    \begin{overpic}[height=\resLen]{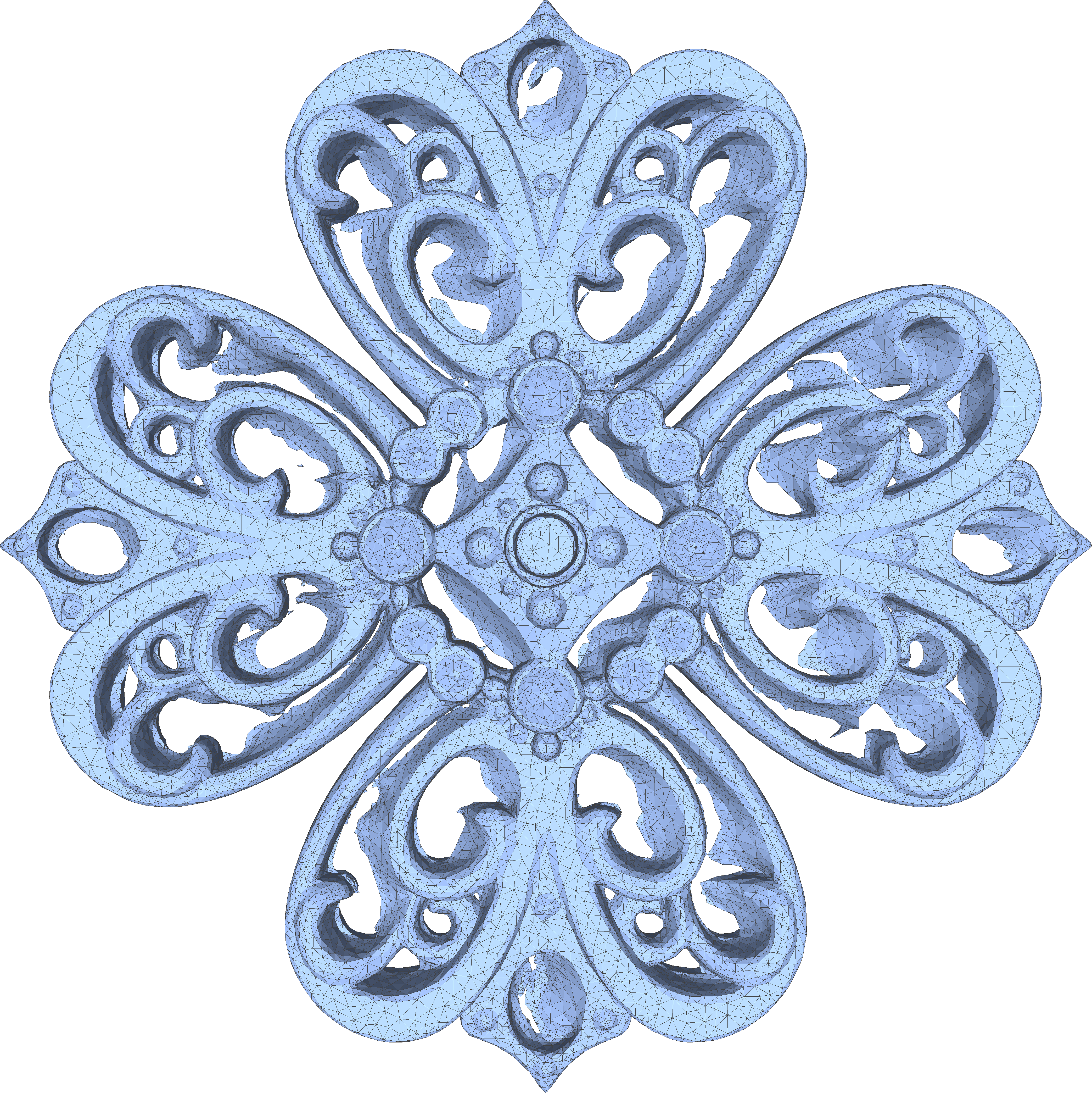}
    \end{overpic} &
    \begin{overpic}[height=\resLen]{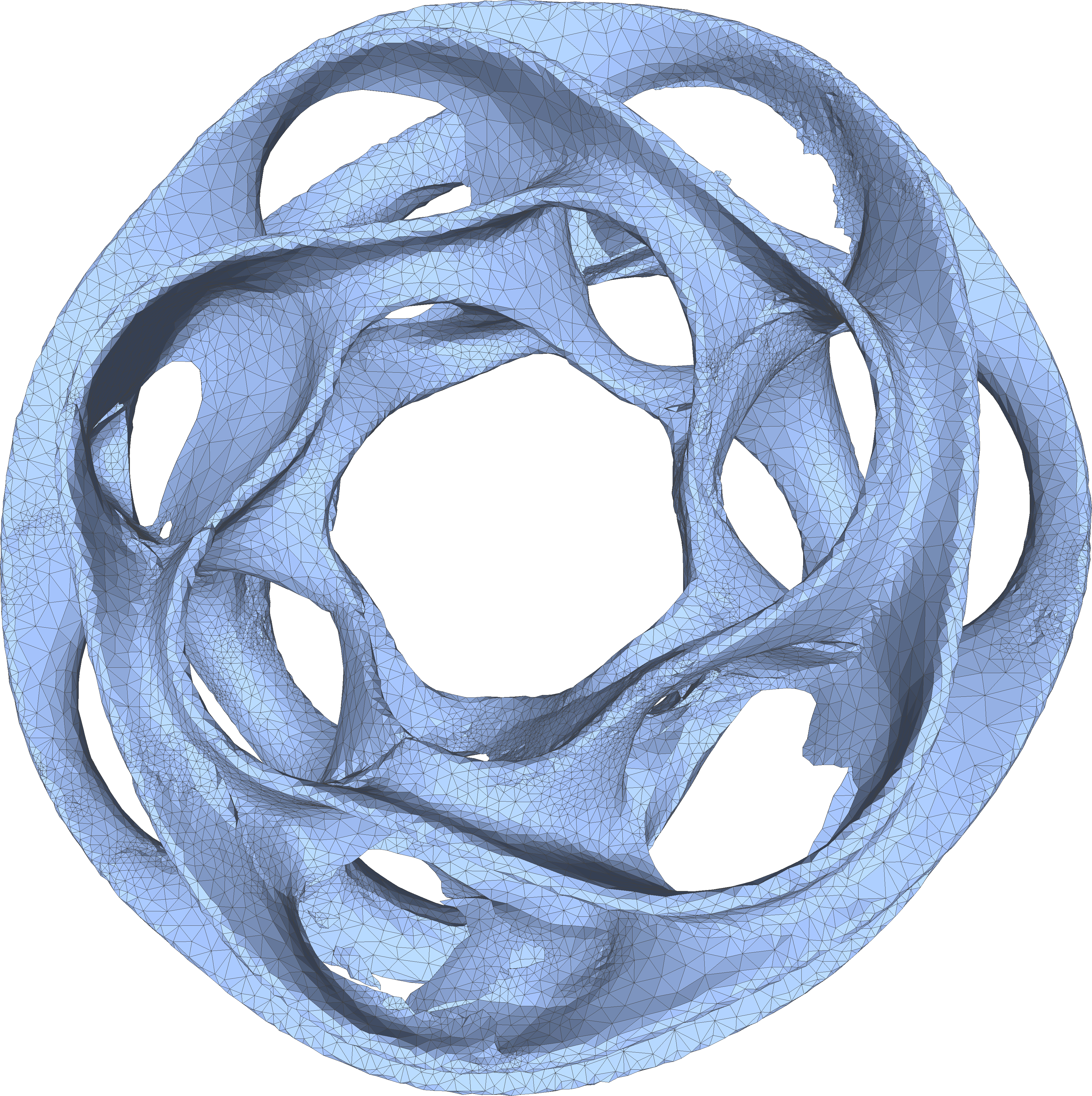}
    \end{overpic} &
    \begin{overpic}[height=\resLen]{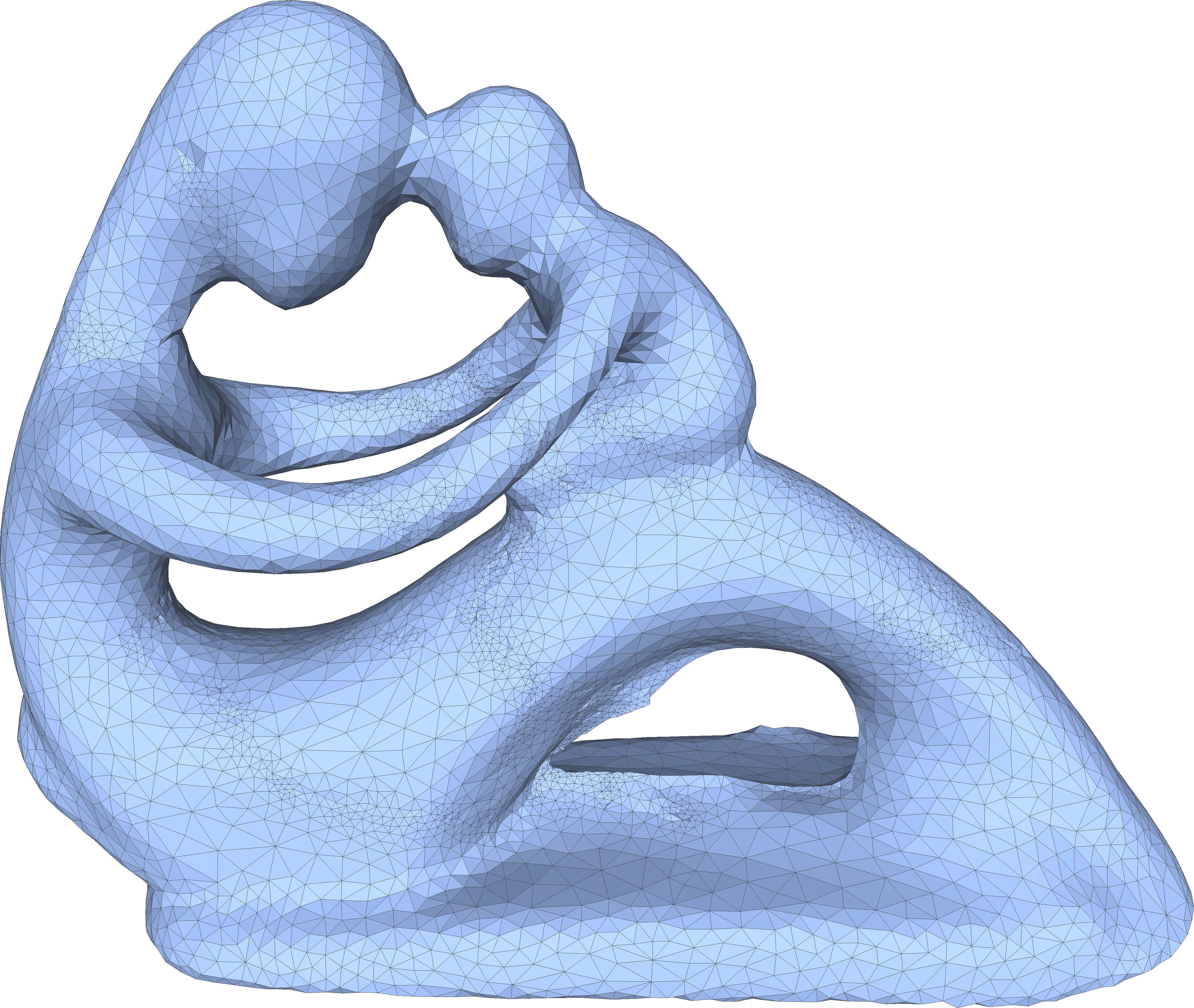}
    \end{overpic}
    \\
     & \raisebox{30pt}{GT} &
    \begin{overpic}[height=\resLen]{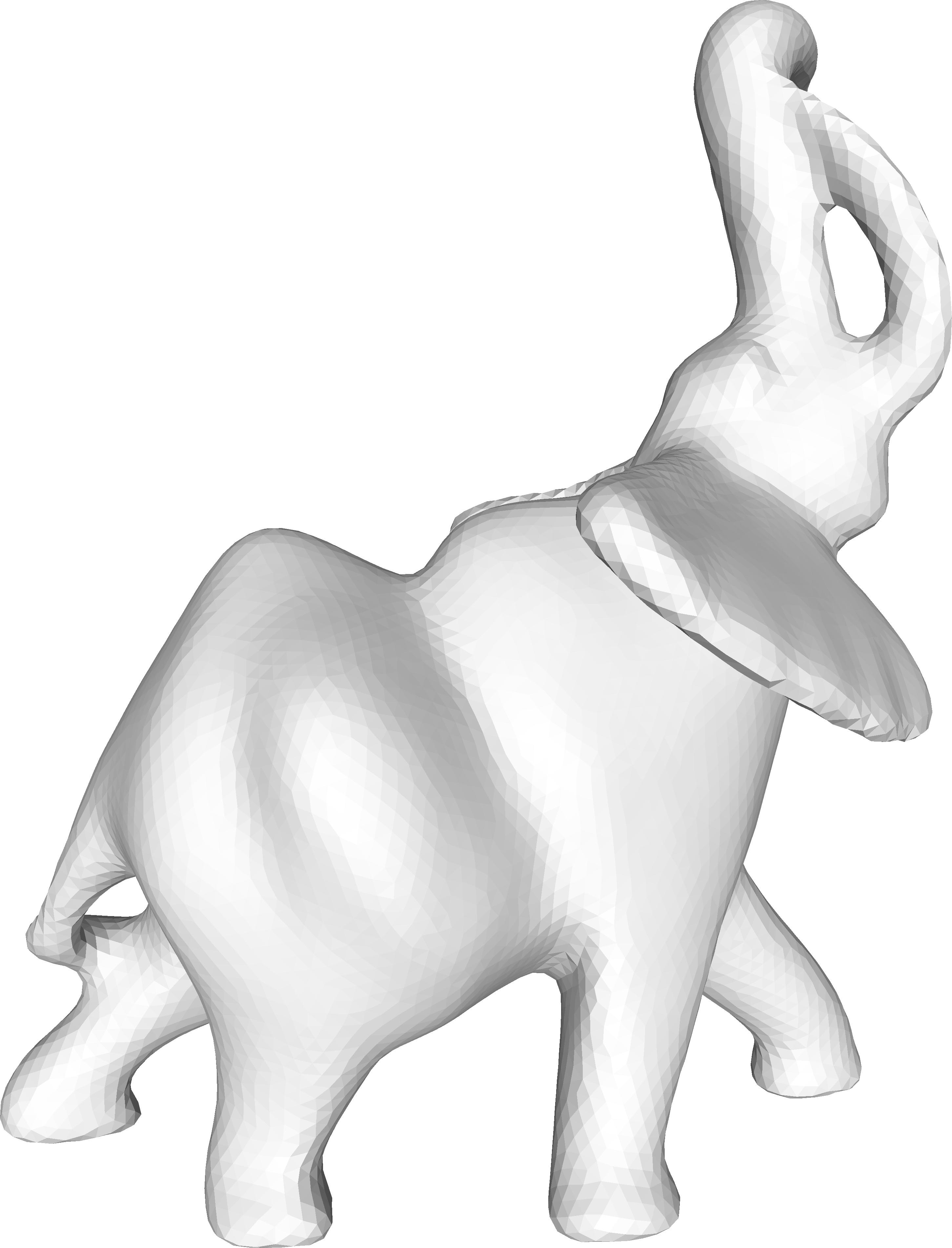}
    \end{overpic} &
    \begin{overpic}[height=\resLen]{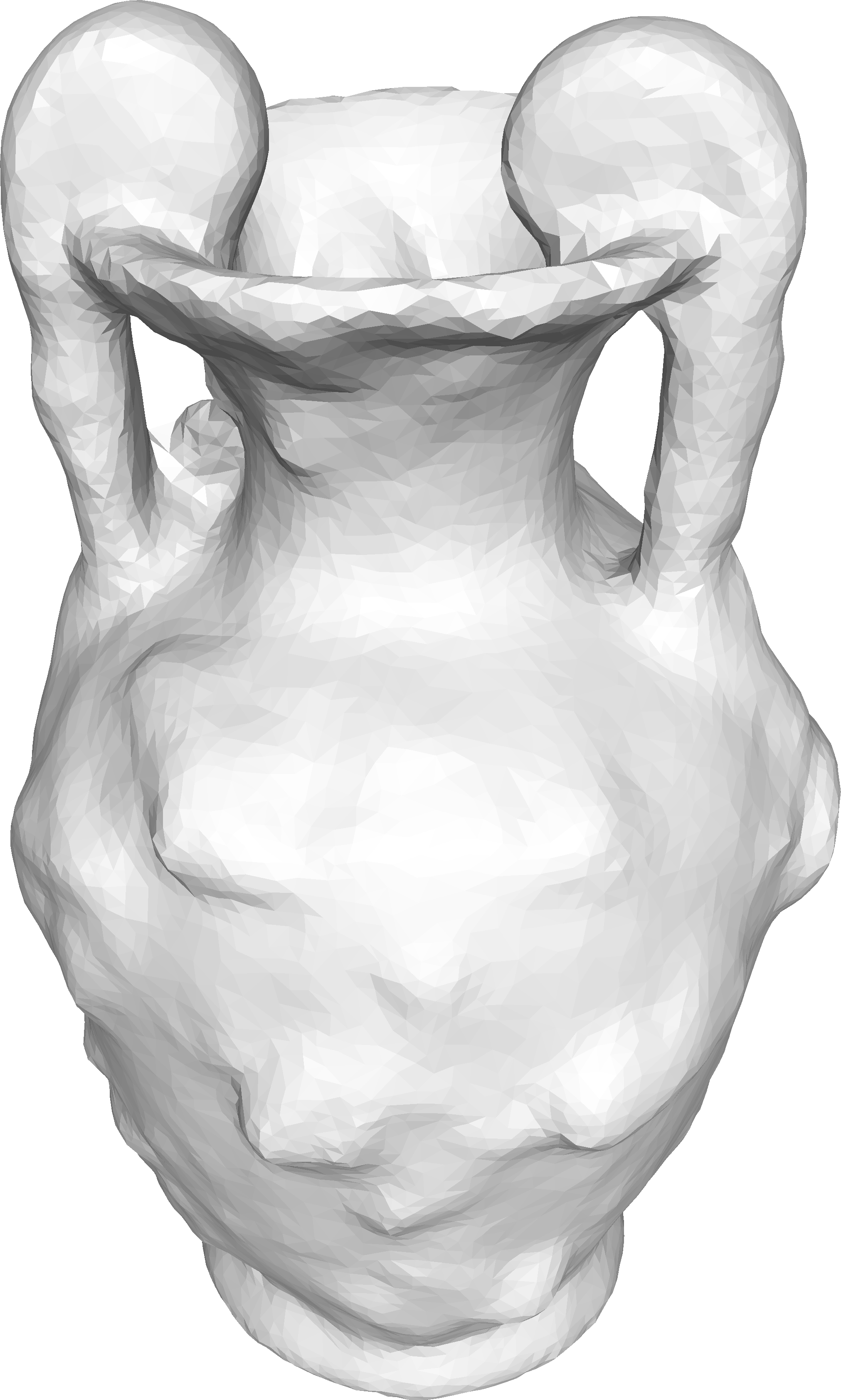}
    \end{overpic} &
    \begin{overpic}[height=\resLen]{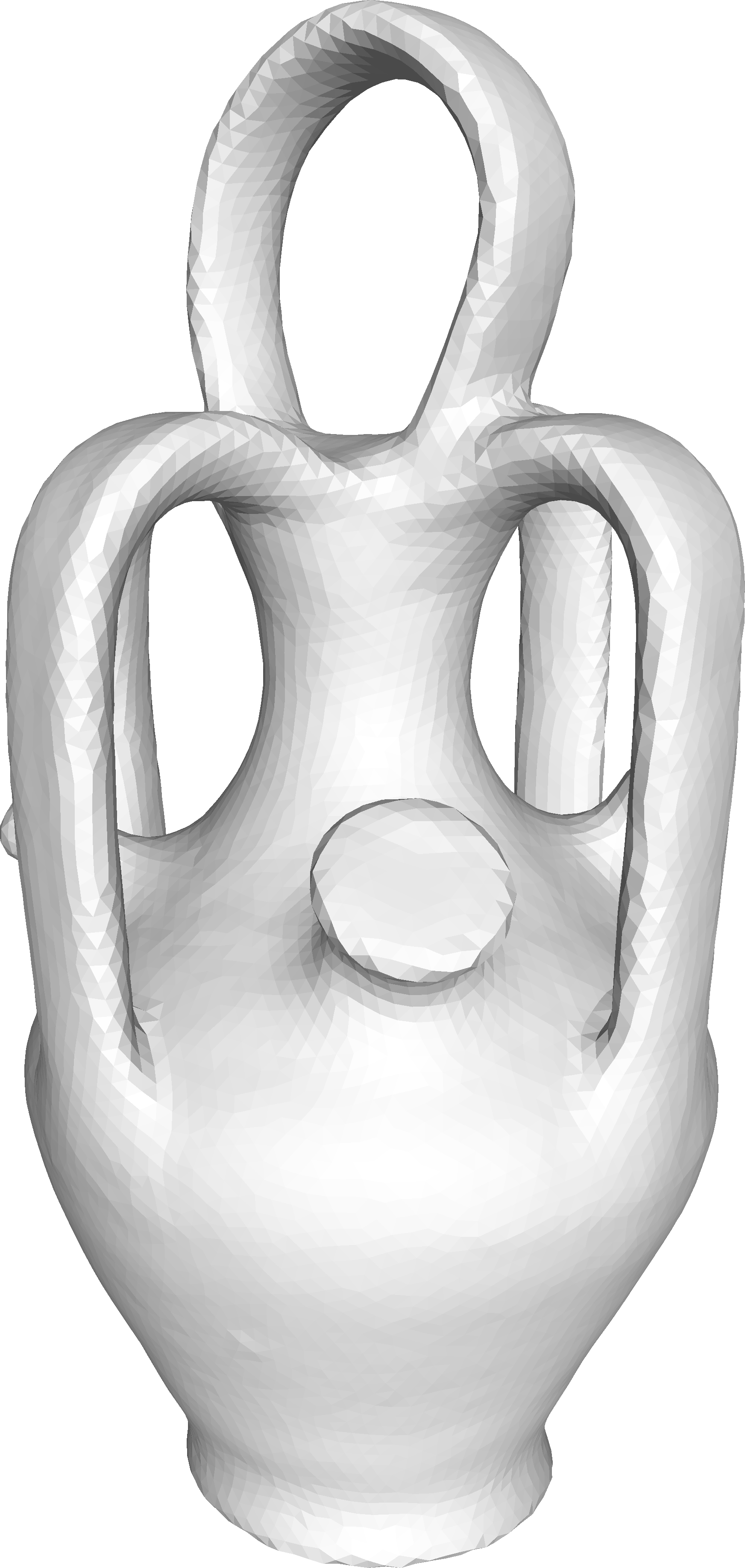}
    \end{overpic} &
    \begin{overpic}[height=\resLen]{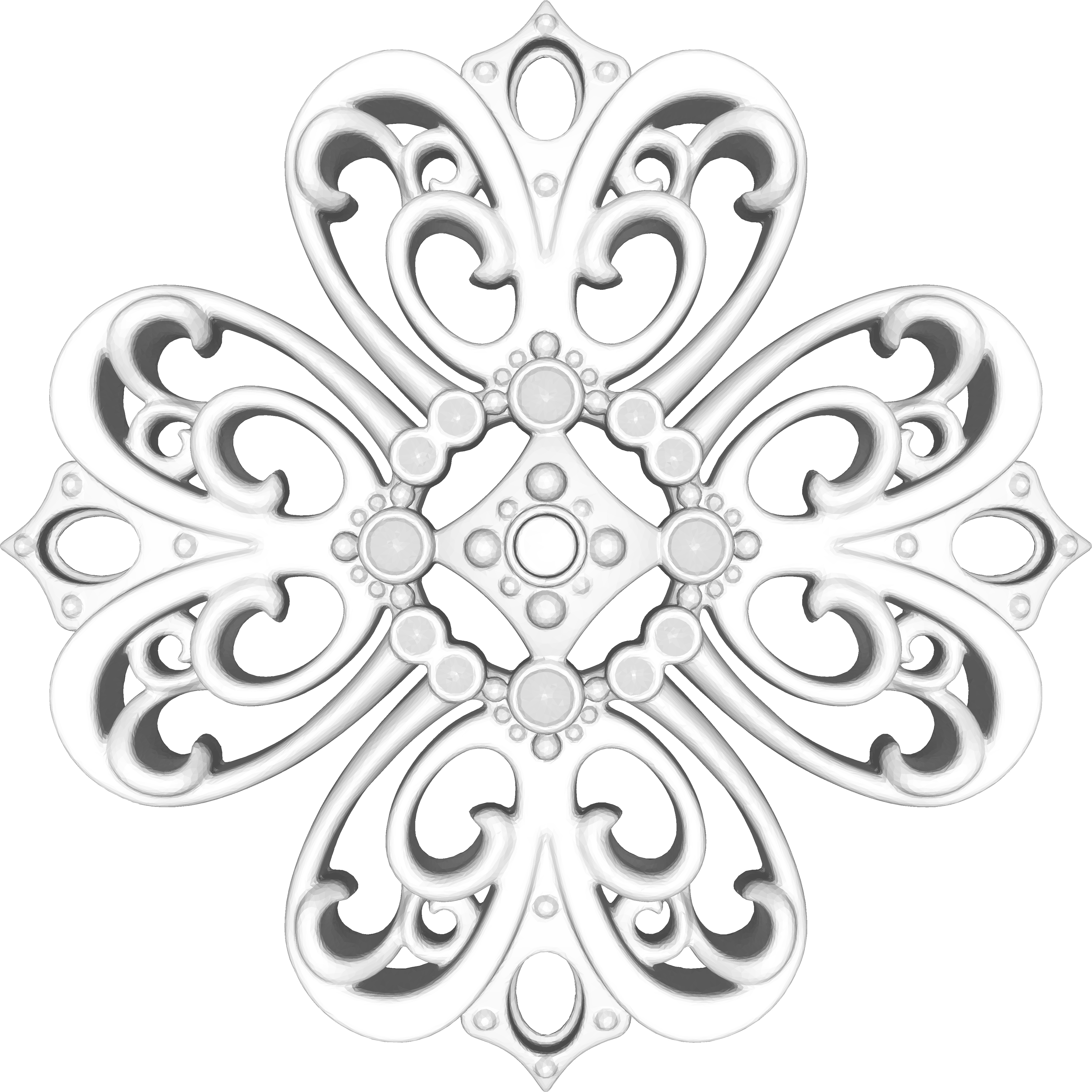}
    \end{overpic} &
    \begin{overpic}[height=\resLen]{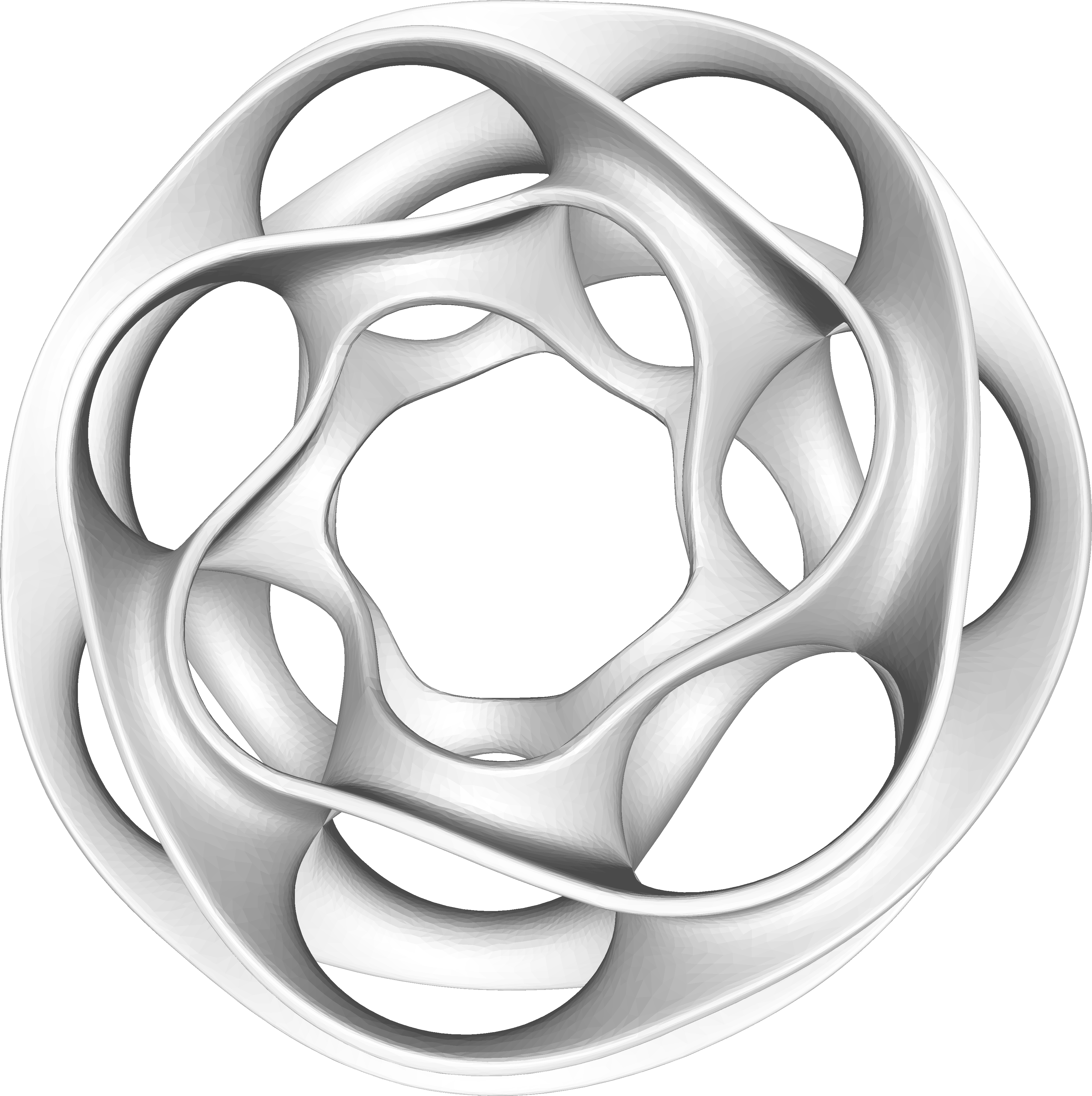}
    \end{overpic} &
    \begin{overpic}[height=\resLen]{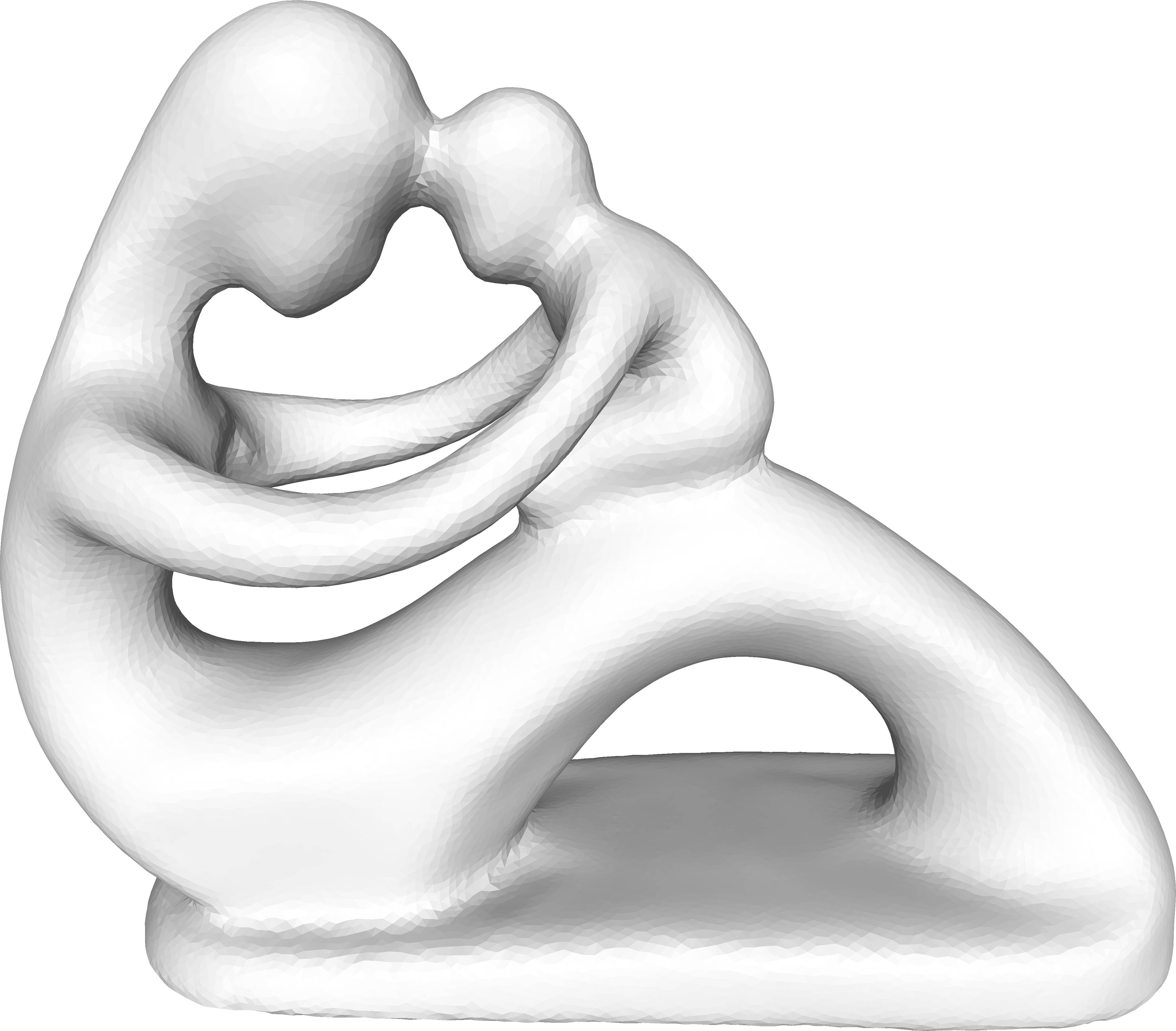}
    \end{overpic} \\
    & & Elephant & Amphora & Botijo & Filigree & Heptoroid & Mother
\end{tabular}

\vspace{1mm}

\captionof{figure}{\textbf{Inverse Rendering for High-Genus Surface Meshes from Multi-View Images.} (Top) Reconstructions using the \textbf{SOTA} method~\cite{Nicolet2021Large}, which \textbf{produces incorrect genus number, leading to incorrect topology}. (Middle) Our method with the \textbf{correct genus number, leading to correct topology}. (Bottom) Challenging High-Genus Ground Truth. Please see \textbf{Appendix \textcolor{red}{D.3}} for quantitative results.}

\label{fig:teaser}

\vspace{4mm}

%% file: sec/0_abstract.tex
\begin{abstract}
We present a topology-informed inverse rendering approach for reconstructing high-genus surface meshes from multi-view images. Compared to 3D representations like voxels and point clouds, mesh-based representations are preferred as they enable the application of differential geometry theory and are optimized for modern graphics pipelines. However, existing inverse rendering methods often fail catastrophically on high-genus surfaces, leading to the loss of key topological features, and tend to oversmooth low-genus surfaces, resulting in the loss of surface details. This failure stems from their overreliance on Adam-based optimizers, which can lead to vanishing and exploding gradients. To overcome these challenges, we introduce an adaptive V-cycle remeshing scheme in conjunction with a re-parametrized Adam optimizer to enhance topological and geometric awareness. By periodically coarsening and refining the deforming mesh, our method informs mesh vertices of their current topology and geometry before optimization, mitigating gradient issues while preserving essential topological features. Additionally, we enforce topological consistency by constructing topological primitives with genus numbers that match those of ground truth using Gauss-Bonnet theorem. Experimental results demonstrate that our inverse rendering approach outperforms the current state-of-the-art method, achieving significant improvements in Chamfer Distance and Volume IoU, particularly for high-genus surfaces, while also enhancing surface details for low-genus surfaces.
\end{abstract}  

%% file: sec/1_intro.tex
\section{Introduction}
\label{sec:intro}
Inverse reconstruction of high-quality 3D surfaces from images is one of the most challenging tasks in computer vision and computer graphics, with diverse applications spanning virtual and augmented reality (VR/AR), medical imaging, robotics, autonomous driving, and 3D printing. Recent advances in reconstruction~\cite{mildenhall2020nerf, kerbl3Dgaussians, Neus, 2dgs24, guo2024tetsphere, hotspot24, Gao2025GraphCutUnwrapping}, generative modeling~\cite{dreamFusion22, liu2023zero1to3, liu2023syncdreamer, long2023wonder3d, Liu2023MeshDiffusion, MvDream23, Wang2025OTTALK}, and inverse rendering~\cite{Nicolet2021Large, Palfinger2022Remesh, mehta2022level, mehta2023topologyderivative, Jung2023, Fu2024BSDF} have led to significant improvements in geometric precision and visual realism, representing a major leap toward real-world deployment of images-driven 3D geometry creation. However, choosing the right 3D representation has remained a critical challenge due to the inherent trade-offs in each 3D representation. For example, voxels~\cite{voxels, VoxNet15, ocnn17, SparseConvNet17} offer structured 3D grids but are computationally expensive at high resolutions. Point clouds~\cite{nichol2022pointe, pointflow, mandikal20183dlmnet, Groueix_2018_CVPR} capture surface points efficiently but lack connectivity, complicating the detection of small holes and other topological features. Triangular and tetrahedral meshes~\cite{meshcnn19, Nicolet2021Large, Jung2023, Palfinger2022Remesh, guo2024tetsphere, guo2024physcomp} provide detailed connected surface representations capturing fine geometric details but rely on complex data structures for efficient local mesh processing~\cite{CCG08, Crane:2013:DGP, botsch2010polygon} due to its irregular structure. Signed Distance Fields (SDFs)~\cite{Vicini2022sdf, neuralflow, deepSDF19, jiang2020sdfdiff, WangSDF2024, mehta2022level, mehta2023topologyderivative} enable smooth and continuous surface representation, but incur high computational cost due to the need for dense 3D grid evaluation and frequent distance queries, and often suffer from loss of surface details during surface mesh extraction via Marching Cubes~\cite{Lorensen1987MarchingCubes}.

Among these 3D representations, meshes are particularly favored in physical simulations and graphics rendering, as modern graphics pipelines are highly optimized for processing \textit{mesh-based} structures efficiently. In engineering, mesh representations are widely used to solve partial differential equations (PDEs). \textit{Mesh-based} finite element simulations have been extensively applied in fields such as solid mechanics \cite{Sifakis2012FEM, liu2018narrow}, fluid mechanics \cite{Bridson2015Fluid, Anderson2020Computational}, aerodynamics~\cite{economon2015su2, Ramamurti2001FlappingAirfoil}, electromagnetics~\cite{Pardo2007hpFEM}, geophysics~\cite{Ren2010AdaptiveFEM, Schwarzbach2011AdaptiveFEM}, and acoustics~\cite{marburg2008computational}, due to their ability to capture intricate geometric details essential for accurate analysis. Furthermore, most mathematical proofs regarding convergence, as well as theoretical analyses of consistency, stability, and error bounds, are specifically developed for \textit{mesh-based} discretizations~\cite{Pasciak1995Review}. Utilizing meshes also allows us to apply theories from discrete differential geometry~\cite{CCG08, Crane:2013:DGP,botsch2010polygon} to inspect surface features such as mean, Gaussian and principal curvatures, enabling the classification of local shapes and the determination of the surface \textit{genus}, a measure of topological complexity that quantities the number of holes or tunnel loops on a surface. Ensuring genus consistency is vital, as mismatches can result in the loss of critical
topological features and lead to visually inaccurate reconstructions. Collectively, these factors establish meshes as the standard representation in both engineering and rendering applications, where 3D surfaces must meet rigorous requirements for accuracy and precision. While converting from other representations to triangular meshes can introduce loss of geometric or topological detail, conversions from meshes typically retain nearly all original information. This inherent adaptability makes \textit{\textit{mesh-based}} representations the preferred choice across a wide range of practical applications.

Recently, there has been a growing shift toward solving inverse rendering problems without relying on neural networks, particularly using \textit{mesh-based} surface representations. Unlike neural-implicit methods such as NeRF~\cite{mildenhall2020nerf, instantNerf22}, neural signed distance fields (SDFs)\cite{deepSDF19, Vicini2022sdf}, and 2D/3D Gaussian Splatting~\cite{kerbl3Dgaussians, 2dgs24}, which are based on the volume rendering equation, \textit{mesh-based} methods~\cite{Nicolet2021Large, Palfinger2022Remesh, Jung2023} adopt differentiable rasterization pipelines to directly optimize surface geometry. This trend has been largely propelled by recent advances in physics-based differentiable rendering (PBDR)~\cite{Loper2014, Loubet2019, Mao18, Zhao2020PBDR, Laine2020diffrast, Mitsuba3, taichiHu19, kato2020differentiablerenderingsurvey} and high-performance GPUs, which enable efficient gradient computation over mesh parameters. Although neural network–free methods~\cite{Nicolet2021Large, Palfinger2022Remesh, Jung2023} are appealing for \textit{real-time} applications, they remain limited in handling complex geometries, particularly high-genus surface meshes. The root cause lies in their heavy reliance on gradient-based optimizers such as Adam~\cite{kingma2017adammethodstochasticoptimization}, which often leads to \textit{vanishing} or exploding gradients. As a result, the geometry becomes overly smoothed, and key topological features are lost. The reconstructed meshes often suffer from \textit{poor} quality, exhibiting artifacts such as topological inconsistency, non-manifold edges, and self-intersections. These issues significantly limit their applicability to downstream tasks such as relighting, physical simulation, and 3D printing. Addressing these limitations is essential for accurately recovering high-quality, high-genus surface from multi-view images.

To overcome these challenges, we introduce a topologyinformed inverse rendering method that leverages an adaptive V-cycle remeshing scheme rooted in geometry processing, in conjunction with an Adam-based optimizer specifically designed for reconstructing high-genus surfaces from
multi-view images. By periodically coarsening and refining
the mesh and subsequently optimizing, our method effectively mitigates gradient issues and preserves key geometric and topological features. We further enforce topological consistency by constructing topological primitives with
genus numbers matching those of the ground truth, utilizing
the Gauss-Bonnet theorem from differential geometry. Experimental results demonstrate that our approach surpasses
previous methods in terms of reconstructed surface quality,
Chamfer Distance, and Volume IoU, particularly for high-genus surface mesh reconstructions.

\noindent
To summarize, we offer \textbf{three principal contributions}:
\begin{itemize}[noitemsep, topsep=0pt, leftmargin=*]
    \item We directly address the core challenges of reconstructing high-genus surface meshes using a \textit{mesh-based} representation within an inverse rendering framework.
    \item We leverage an adaptive V-cycle remeshing scheme in conjunction with an Adam-based optimizer to effectively mitigate gradient issues, enhancing topological awareness, preserving key topological features, and enforcing topological consistency by establishing a homeomorphism between initial and ground truth surfaces, achieving genus invariance.
    \item We demonstrate superior performance in reconstructing high-genus surfaces, evaluated both qualitatively and quantitatively using Chamfer Distance and Volume IoU.
\end{itemize}

%% file: sec/2_related.tex
\section{Related Work}


\noindent\textbf{3D Reconstruction.} Reconstructing 3D objects from multi-view images is inherently an ill-posed problem. As a result, extensive research has been devoted to addressing it. Early neural network-based methods typically rely on 2D image encoders and 3D decoders trained on 3D datasets, using both explicit and implicit representations, including voxels~\cite{Chen_2019_CVPR, xie2019pix2vox, Xie2020} and point clouds~\cite{mandikal20183dlmnet, Groueix_2018_CVPR}. Although promising, these approaches face challenges due to their underlying 3D representations. The former requires high 3D grid resolution to capture fine surface details, leading to significant computational overhead, while the latter relies on dense points to capture topological features such as small holes or tunnels. Furthermore, both voxels and point clouds require post-processing to convert to meshes for applications such as relighting and physical simulation. Recent advancements in implicit representations, particularly neural radiance fields (NeRF)~\cite{mildenhall2020nerf}, have led to generative models based on neural fields~\cite{Chan2021, chan2021pigan}, enabling 3D shape learning from 2D images via differentiable rendering. However, converting implicit representations to explicit surfaces often leads to loss of fine geometric detail. Recent methods such as 3D Gaussian Splatting (GS)~\cite{kerbl3Dgaussians} and TetSphere Splatting~\cite{guo2024tetsphere} have shifted focus back to explicit representations, leveraging Gaussian kernels and tetrahedral meshes, respectively, to improve both visual and reconstruction quality. Tetrahedral meshes enable lossless surface extraction but are more memory-intensive than surface meshes, while 3D Gaussian Splats require post-processing, which may result in loss of surface details. For applications such as relighting, physical simulation, and 3D printing, directly using 3D surface \textit{mesh-based} representation without sacrificing detail is highly desirable. 

\noindent\textbf{Inverse Rendering for 3D.} Recent advancements in physics-based differentiable rendering (PBDR)~\cite{Zhao2020PBDR, Laine2020diffrast, Mitsuba3, SVBRDFRecovery21}, largely enabled by high-performance NVIDIA GPUs, have significantly expanded the capabilities of inverse rendering, allowing efficient gradient-based optimization to recover scene parameters such as geometry, texture, and lighting from multi-view images. With Adam-based solvers, recent works~\cite{Nicolet2021Large, Jung2023} have demonstrated the extraction of high-quality 3D triangular meshes in seconds to minutes under consistent lighting, \textbf{notably without relying on neural networks}. To handle more complex environmental conditions, approaches such as~\cite{Munkberg_2022Env, neuralflow} incorporate neural networks to indirectly generate triangular meshes, textures, and lighting, though these methods incur significantly higher computational cost. In addition, methods like~\cite{mehta2022level, mehta2023topologyderivative} formulate inverse rendering using level set methods to represent surfaces implicitly. These methods are computationally expensive but particularly suited for high-genus topology due to its powerful 3D representation. In contrast,~\cite{SVBRDFRecovery21} apply inverse rendering to simultaneously recover both geometry and appearance for high-genus surfaces, such as genus-1 shapes. However, reconstructing high-genus surface meshes using \textit{mesh-based} representations without neural networks remains challenging, as most existing methods are not topologically aware. Sole reliance on Adam optimization often produces overly smooth surfaces and suffers from vanishing or exploding gradients, leading to the loss of critical geometric and topological features. Addressing these limitations is essential to advance \textit{mesh-based} inverse rendering that robustly captures high-genus topology and fine surface detail.

%% file: sec/3_background.tex
\section{Theoretical Foundations}
\noindent\textbf{Differential Geometry of Surfaces.} Differential geometry provides the mathematical foundation to analyze the curvatures of the surface, which is often represented by a vector-valued function:

\begin{equation}
\vec{\mathbf{r}}(u, v) = \left(x_1(u, v), x_2(u, v), x_3(u, v) \right) \in \mathbb{R}^3,
\label{eq:surface}
\end{equation}
where \( (u, v) \in \mathbb{R}^2 \) are the coordinate parameters of the surface. At each point on the surface, the vectors \(\vec{\mathbf{r}}_u = \frac{\partial \vec{\mathbf{r}}}{\partial u}\) and \(\vec{\mathbf{r}}_v = \frac{\partial \vec{\mathbf{r}}}{\partial v}\) form a local basis for the tangent plane.

\noindent\textbf{Principal Curvatures.}
One of the key reasons our adaptive V-Cycle remeshing can be topology-informed and preserves topological and geometric features is that it is guided by principal curvatures \( k_1 \) and \( k_2 \in \mathbb{R} \), which are defined as:

\begin{equation}
k_1 = \max_{\phi} k_{\vec{\mathbf{n}}}(\phi), \quad k_2 = \min_{\phi} k_{\vec{\mathbf{n}}}(\phi),
\label{eq:principal_curvatures}
\end{equation} 
where \( k_{\vec{\mathbf{n}}}(\phi) \) represents the normal curvature in the direction of the polar angle \( \phi \in [0, 2\pi) \).

\noindent\textbf{Mean and Gaussian Curvatures.}
In differential geometry, both the mean curvature and Gaussian curvature at a point on the surface are deeply linked to the principal curvatures. For instance, the mean curvature \( H \in \mathbb{R} \) at a point on a surface is defined as the average of the principal curvatures \( k_1 \) and \( k_2 \):
\begin{equation}
H = \frac{1}{2\pi} \int_0^{2\pi} k_{\vec{\mathbf{n}}}(\phi) \, d\phi = \frac{k_1 + k_2}{2}.
\label{eq:mean_curvature}
\end{equation}
Similarly, the Gaussian curvature \( K \in \mathbb{R} \) at the same point is defined as the product of the principal curvatures  \( k_1 \) and \( k_2 \):
\begin{equation}
K = k_1 \cdot k_2.
\label{eq:gaussian_curvature}
\end{equation} 
Assuming we know both \( H \) and \( K \) from Equations~\ref{eq:mean_curvature} and \ref{eq:gaussian_curvature}, we can simply solve for the principal curvatures \( k_1 \) and \( k_2 \), yielding:
\begin{equation}
k_1, k_2 = H \pm \sqrt{H^2 - K}.
\label{eq:principal_curvatures_from_H_K}
\end{equation} In practice, we always triangulate the surface into a mesh, which allows us to compute the discrete mean and Gaussian curvatures, enabling us to approximate the principal curvatures. We will see this in the next few sections.

\noindent\textbf{Discrete Gaussian Curvature.}
Working on the same mesh, the discrete Gaussian curvature \( K_i \) at vertex \( v_i \) is given by the angle deficit method \cite{CCG08, botsch2010polygon}:

\begin{equation}
\mathbf{K_i} = 
\begin{cases} 
2\pi - \sum_{j \in N(i)} \theta_j, & \text{if } v_i \notin B, \\
\pi - \sum_{j \in N(i)} \theta_j, & \text{if } v_i \in B,
\end{cases}
\label{eq:discrete_gaussian_curvature}
\end{equation} where \( \theta_j \) are the interior angles at \( v_i \) across adjacent triangles, \( N(i) \) is the one-ring neighborhood of faces around \( v_i \), and \( B \) denotes boundary vertices. 

\noindent\textbf{Discrete Mean Curvature.}  
Interestingly, by leveraging the discrete Laplacian-Beltrami operator \( \mathbf{L} \)  \cite{Nealen2006, botsch2004remeshing}, we can approximate the mean curvature \( H_i \) at each vertex \( v_i \) as:

\begin{equation}
H_i = \frac{1}{2} \| \mathbf{L} \vec{\mathbf{x}}_i \|,
\label{eq:discrete_mean_curvature}
\end{equation} 
where \( \mathbf{L} \vec{\mathbf{x}}_i = \sum_{j \in N(i)} w_{ij} (\vec{\mathbf{x}}_j - \vec{\mathbf{x}}_i) \) is the discrete Laplacian applied to the position \( \vec{\mathbf{x}}_i \) of \( v_i \) using either uniform or cotangent weight with the factor \( \frac{1}{2} \) included to align with the continuous definition. 
Hence, we can approximate the principal curvatures using the discrete Gaussian curvature and mean curvatures using earlier Equation \ref{eq:principal_curvatures_from_H_K}.

\noindent\textbf{Gauss-Bonnet Theorem.}
In order to enforce topological consistency as to prevent topological mismatch between initial and ground truth surfaces, we apply the Gauss-Bonnet theorem which links the genus \( g \) of a surface \( S \) to its Euler characteristic \( \chi(S) \) via:

\input{fig/mismatch}

\input{fig/pipeline}

\begin{equation}
\int_S K \, dA + \int_{\partial S} k_g \, ds = 2\pi \chi(S),
\label{eq:gauss_bonnet}
\end{equation} 
where \( K \) denotes Gaussian curvature, \( k_g \) denotes the geodesic curvature along the boundary \( \partial S \), and \( \chi(S) = 2 - 2g \) for a closed, orientable surface without boundary. For a triangular mesh \( \mathcal{M}\), Euler characteristic is defined as:

\begin{equation}
\chi(S) = |V| + |F| - |E|,
\label{eq:euler_characteristic}
\end{equation} where \( |V| \), \( |E| \), and \( |F| \) represent the number of vertices, edges, and faces, respectively. Consequently, the genus \( g \) of a closed, orientable surface can be computed as:

\begin{equation}
g = 1 - \frac{|V| + |F| - |E|}{2}.
\label{eq:genus_formula}
\end{equation} By applying Equations~\ref{eq:genus_formula}, we can compute the genus \( g \) of any triangulated 3D surface, enabling us to establish a homeomorphism between initial triangulated surfaces and the ground truth ones to circumvent genus mismatch, as shown in Figure~\ref{fig:mismatch}. For an in-depth discussion on discrete differential geometry, we recommend \cite{botsch2010polygon, CCG08, Crane:2013:DGP}.

%% file: fig/mismatch.tex
\begin{figure}[t]
    \centering
    \includegraphics[width=0.9\linewidth]{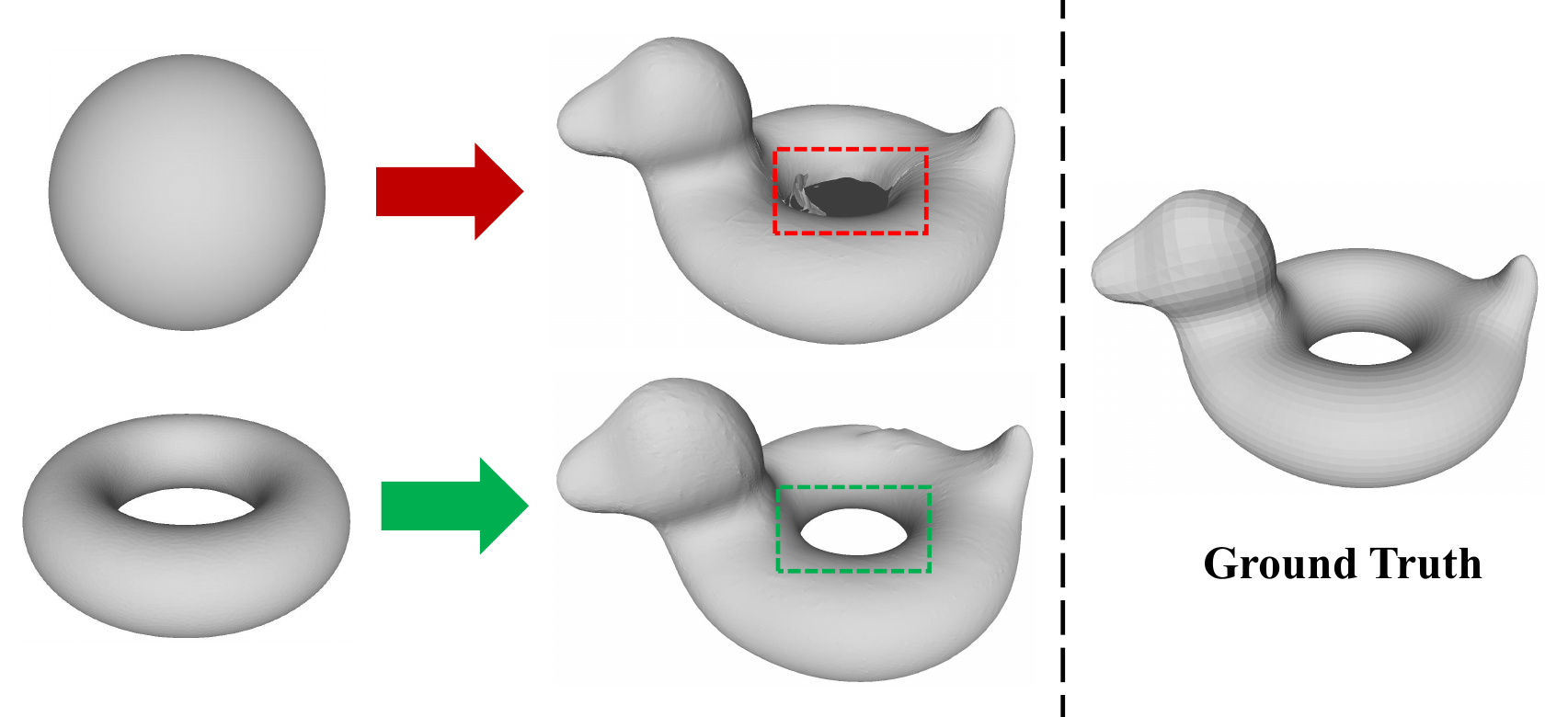}
       \caption{Topological Consistency: The top row shows a sphere with \textbf{genus 0}, not homeomorphic to the Bob surface's \textbf{genus 1} ground truth, resulting in topological inconsistency. The bottom row shows a torus with \textbf{genus 1}, ensuring topological consistency.}
    \label{fig:mismatch}
\end{figure}

%% file: fig/pipeline.tex
\begin{figure*}[t]
    \centering
    \includegraphics[width=\textwidth]{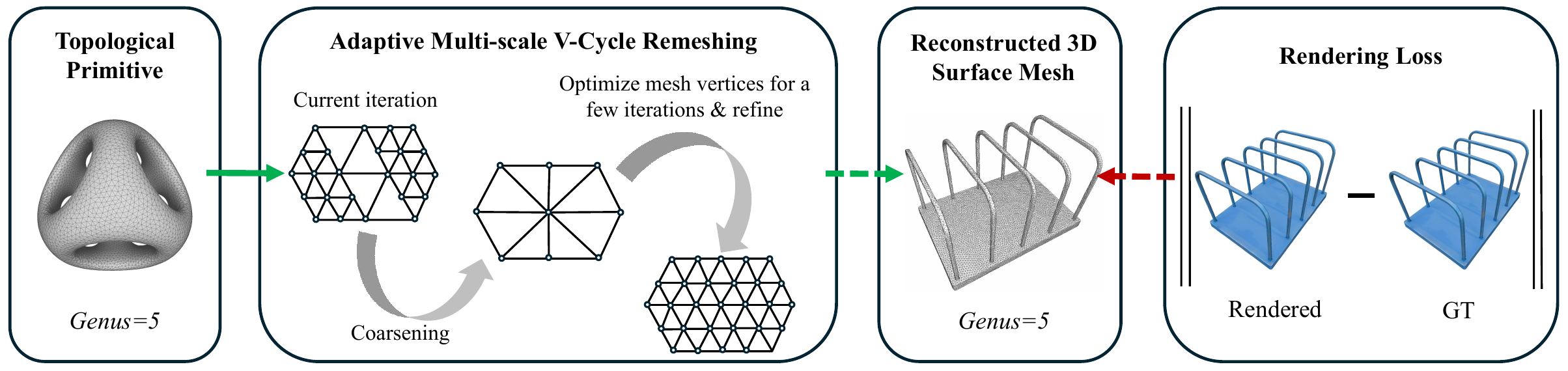}
    \caption{\textbf{Overall Pipeline}: A triangulated topological primitive, with genus matching that of the ground truth, undergoes adaptive V-cycle remeshing with periodic coarsening and refining stages, followed by optimization using an Adam-based optimizer to minimize the multiview rendering loss.}

    \label{fig:pipeline}
\end{figure*}

%% file: sec/4_method.tex
\section{Method}
\subsection{Problem Formulation}
We formulate the inverse rendering problem for high genus surfaces as:

\begin{equation} 
\begin{aligned}
   \arg\min_{\mathbf{x}} \quad & \Phi(R(\mathbf{x})) + \mathbf{w_1}\operatorname{tr}(\mathbf{x}^T \mathbf{L} \mathbf{x}) \\
    \text{s.t.} \quad & \det(\mathbf{J}_R^{(k)}) > 0, \quad \forall k \in \{1, \dots, |F|\}
\end{aligned}
\label{eq:constrained}
\end{equation} where \( \mathbf{x} \in \mathbb{R}^{n \times 3} \) represents the mesh vertex positions, \( \mathbf{L} \in \mathbb{R}^{n \times n} \) is the uniform bi-Laplacian matrix \cite{Botsch2004a, Jacobson2010}, \( R(\cdot) \) denotes the rendering function, \( \Phi(\cdot) \) quantifies the rendering loss between rendered and target images, \( \operatorname{tr}(\cdot) \) is the sum of the diagonal entires of a matrix,  \( |F| \) is the number of faces of the triangular mesh, and \( \mathbf{J}_R^{(k)} \) represents the Jacobian for each triangle \( k \). Reformulating the Equation~\ref{eq:constrained} as an unconstrained optimization problem simply yields:

\begin{equation} 
\begin{aligned}
    \arg\min_{\mathbf{x}} \, &\Phi(R(\mathbf{x})) + \mathbf{w_1} \operatorname{tr}(\mathbf{x}^T \mathbf{L} \mathbf{x}) \\
    &+ \mathbf{w_2} \sum_{k=1}^{|F|} \left( \min \{ 0, \det(\mathbf{J}_R^{(k)}) \} \right)^2
\end{aligned}
\label{eq:unconstrained}
\end{equation} allowing for optimization via gradient descent solvers. In this formulation, a large \( \mathbf{w_1} \) enforces mesh smoothness, while a large \( \mathbf{w_2} \) prevents triangle inversion.


\input{fig/mesh_operate}

\input{fig/qual_highgenus}

\subsection{Adaptive V-Cycle Remeshing}
Our method leverages adaptive remeshing guided by the principal curvatures \( k_1 \) and \( k_2 \), derived from Equation~\ref{eq:principal_curvatures_from_H_K} and approximated using Equations~\ref{eq:discrete_gaussian_curvature} and \ref{eq:discrete_mean_curvature} . \textbf{The overall pipeline} is illustrated in Figure~\ref{fig:pipeline}. To maintain an optimal mesh structure that balances computational efficiency and topological awareness, periodic coarsening and refining are essential. Coarsening alone reduces computational load by simplifying the mesh but risks losing critical surface details, while refining enhances detail at the cost of increased memory usage. By alternating coarsening and refining, we reposition vertices \( \mathbf{x} \in \mathbb{R}^{n \times 3} \) to create a high-quality mesh that accurately captures geometric features while maintaining topological consistency. After each remeshing step, vertex positions are further optimized using a re-parametrized Adam-based optimizer \cite{Nicolet2021Large} to minimize rendering loss. This iterative process ensures that the surface mesh is informed by topological features for accurate reconstruction of high-genus surfaces while enhancing surface details for low-genus surfaces. \textbf{See Appendix \textcolor{red}{C.2} for the adaptive V-Cycle remeshing pseudo-code.}

More specifically, our adaptive remeshing algorithm, leverages the half-edge data structure \cite{CCG08, botsch2010polygon, botsch2004remeshing} for coarsening and refinement, proceeds as follows:

\begin{enumerate}[noitemsep, topsep=0pt, leftmargin=*]
    \item \textbf{Edge Splitting:} We split edges with high curvature (Figure~\ref{fig:mesh_operations}a) to enhance resolution in regions with sharp surface details while uniformly splitting low-curvature edges based on the averaged edge length.
    \item \textbf{Edge Collapsing:} We collapse short edges (Figure~\ref{fig:mesh_operations}b) until all edge lengths meet a minimum threshold, optimizing mesh structure by reducing excess vertices where high resolution is unnecessary.
    \item \textbf{Edge Flipping:} We flip edges (Figure~\ref{fig:mesh_operations}c) to improve vertex valence (Targeting a valence of 6 for any closed surface), maintaining mesh stability and uniformity.
    \item \textbf{Tangential Smoothing:} We smooth vertices in direction parallel to the tangent plane to improve triangle quality.
\end{enumerate}

%% file: fig/mesh_operate.tex
\begin{figure}[t]
    \centering
    \setlength{\resLen}{0.45\linewidth}
    \addtolength{\tabcolsep}{0pt}
    \hskip 0pt
    \begin{tabular}{ccc} 
        \begin{overpic}[height=\resLen]{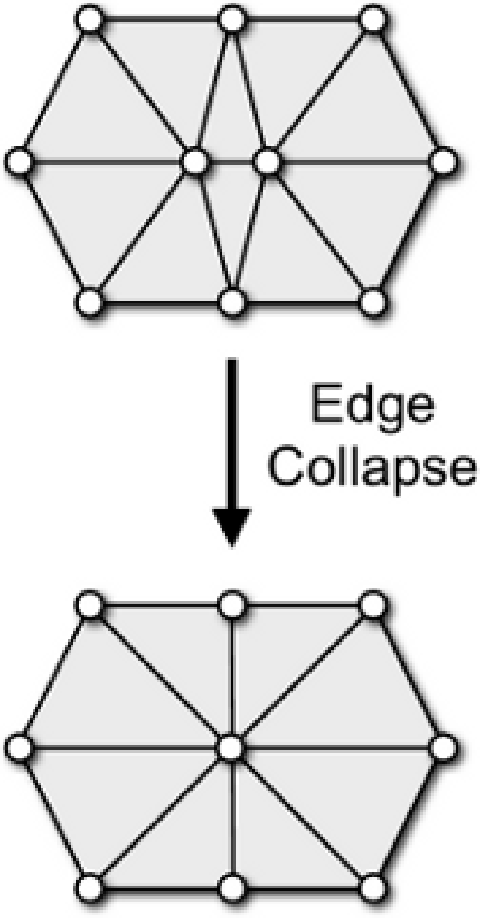}
        \end{overpic} &
        \begin{overpic}[height=\resLen]{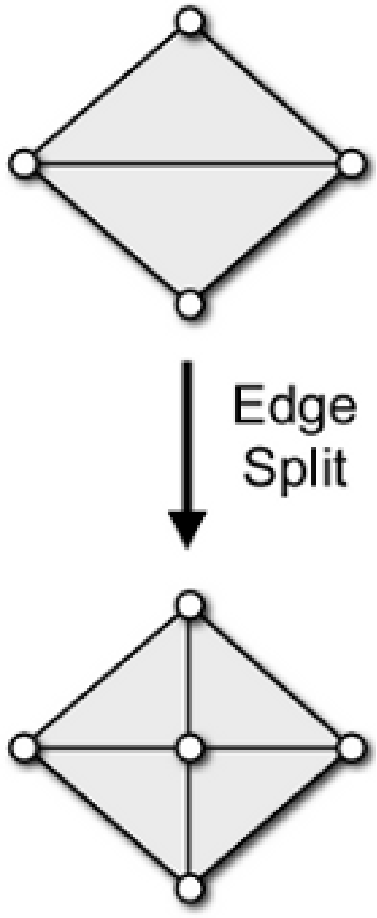}
        \end{overpic} &
        \begin{overpic}[height=\resLen]{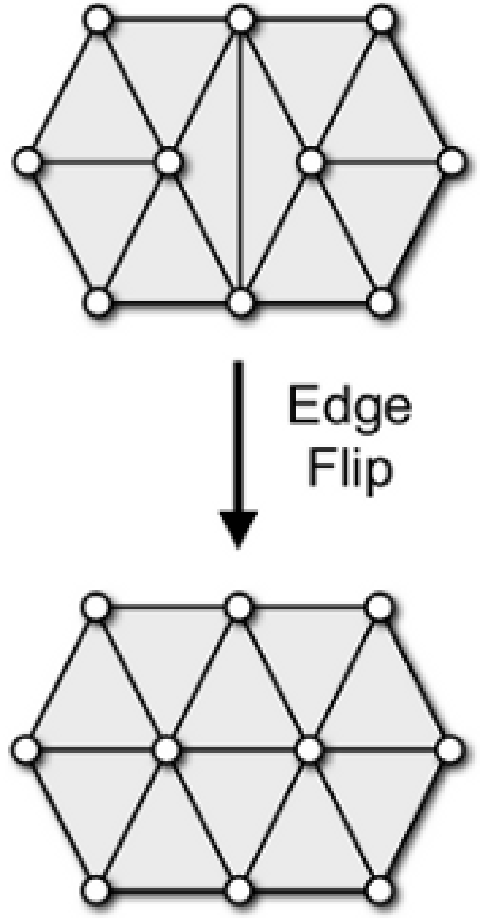}
        \end{overpic}
        \\
        (a) & (b) & (c)        
    \end{tabular}
    \caption{\textbf{Visualization of topology-preserving local mesh operations}~\cite{meshFigs23}. (a) Edge collapse, (b) Edge split, and (c) Edge flip.}
    \label{fig:mesh_operations}
\end{figure}

%% file: fig/qual_highgenus.tex
\begin{figure*}[t]
    \centering
    \setlength{\resLen}{0.12\linewidth}
    \addtolength{\tabcolsep}{-4pt}
    \hskip -5pt
    \begin{tabular}{ccccc@{\hskip 8pt}c} 
        \raisebox{20pt}{\thead{Nicolet et al.\\\cite{Nicolet2021Large}}} &
        \begin{overpic}[height=\resLen]{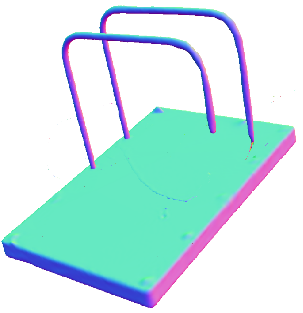}
        \end{overpic} &
        \begin{overpic}[height=\resLen]{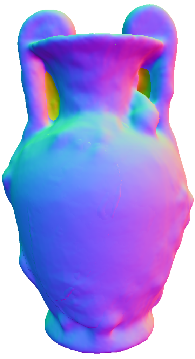}
        \end{overpic} &
        \begin{overpic}[height=\resLen]{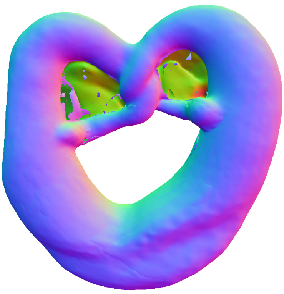}
        \end{overpic} &
        \begin{overpic}[height=\resLen]{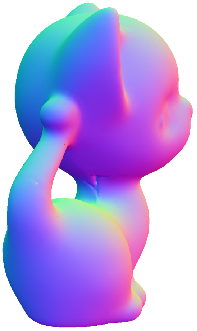}
            \put(0,0){
                \color{red}\linethickness{0.1pt}\polygon(17,36)(27,36)(27,48)(17,48)}
            \put(45,0){
                \color{red}\linethickness{1pt}\frame{
                    \includegraphics[clip, trim=30 55 40 75, height=0.3\resLen]{img/genus_0/kitten/nico_1.png}}}
        \end{overpic} &
        \begin{overpic}[height=\resLen]{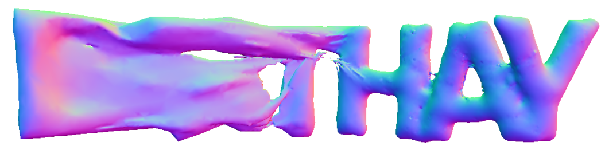}
        \end{overpic} 
        \\
        \raisebox{20pt}{Ours} &
        \begin{overpic}[height=\resLen]{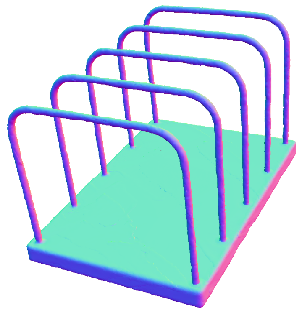}
        \end{overpic} &
        \begin{overpic}[height=\resLen]{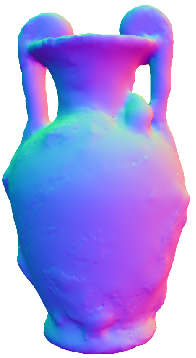}
        \end{overpic} &
        \begin{overpic}[height=\resLen]{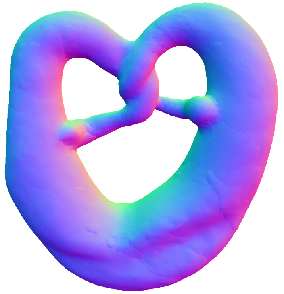}
        \end{overpic} &
        \begin{overpic}[height=\resLen]{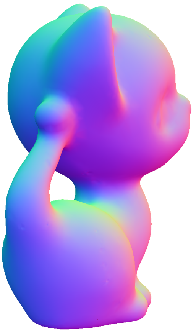}
            \put(0,0){
                \color{red}\linethickness{0.1pt}\polygon(17,36)(27,36)(27,48)(17,48)}
            \put(45,0){
                \color{red}\linethickness{1pt}\frame{
                    \includegraphics[clip, trim=30 55 40 75, height=0.3\resLen]{img/genus_0/kitten/ours_1.png}}}
        \end{overpic} &
        \begin{overpic}[height=\resLen]{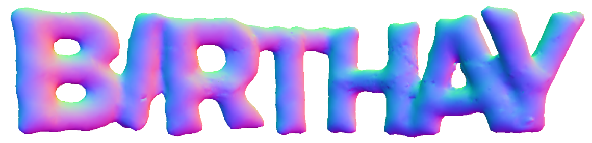}
        \end{overpic} 
        \\
        \raisebox{20pt}{GT} &
        \begin{overpic}[height=\resLen]{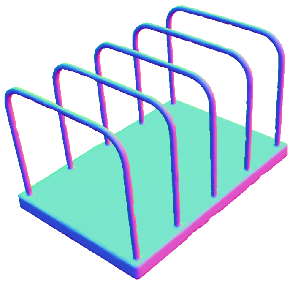}
        \end{overpic} &
        \begin{overpic}[height=\resLen]{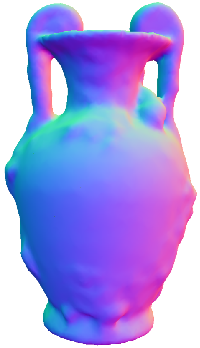}
        \end{overpic} &
        \begin{overpic}[height=\resLen]{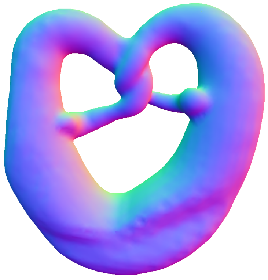}
        \end{overpic} &
        \begin{overpic}[height=\resLen]{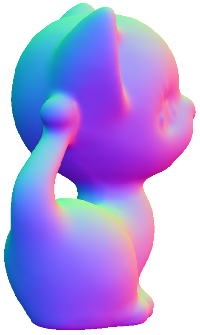}
            \put(0,0){
                \color{red}\linethickness{0.1pt}\polygon(17,36)(27,36)(27,48)(17,48)}
            \put(45,0){
                \color{red}\linethickness{1pt}\frame{
                    \includegraphics[clip, trim=30 55 40 75, height=0.3\resLen]{img/genus_0/kitten/gt.png}}}
        \end{overpic} &
        \begin{overpic}[height=\resLen]{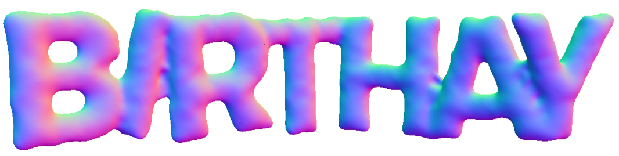}
        \end{overpic} \\
        & Sorter & Amphora & Pretzel & Kitten & Birthday
    \end{tabular}
    \caption{Qualitative High-Genus Reconstruction: Rendered Views in Normal Maps Using Topologically Consistent Triangulated Primitives \textbf{(Genus 1, 2, 3, 4 and 5)}. Please see \textbf{Appendix \textcolor{red}{A.2}} for the complete set of multi-view high-genus surface qualitative results. }

    \label{fig:qualitative_highgenus}
\end{figure*}

%% file: sec/5_results.tex
\section{Experiments and Results}

\input{fig/genus0}

\noindent\textbf{Implementation details}
Our method leverages an adaptive remeshing algorithm developed in C++ using a half-edge-based data structure \cite{CCG08, botsch2010polygon, botsch2004remeshing} to enable efficient local mesh operations. To balance computational efficiency and surface quality on a single 24GB RTX 3090, we set a frequency range of 130–200 iterations to periodically coarsen or refine the deforming mesh vertices \( \mathbf{x} \in \mathbb{R}^{n \times 3} \); these updated mesh vertices are then passed to the current optimization loop, where they are rendered and optimized according to the \( l_1 \) rendering loss \( \Phi(R(\mathbf{x})) \), implemented as a PyTorch extension using CUDA. To ensure timely reconstruction, we limit the optimization to 1500 iterations for low-genus surfaces and 3000 iterations for high-genus surfaces in both the baseline method and our approach. For multi-view reconstruction, we generate RGBA images of the ground truth surface by uniformly capturing \cite{extremelearning_sphere} multi-view images with a radial distance away from a unit sphere to serve as our ground truth images, using a batch of 36 at a resolution of \( 1024 \times 1024 \) for both low- and high-genus surfaces, and a batch of 120--360 at \( 256 \times 256 \) in rare cases where the rendered views fail to capture essential topological features. Finally, we render the deforming mesh with vertices \( \mathbf{x} \) using the same camera poses as those of the ground truth, allowing us to compute the rendering loss effectively. With this setup, the average reconstruction time across low-genus models is approximately 1 minute, while for high-genus models, it is approximately 2–5 minutes on a \textbf{single} 24GB NVIDIA RTX 3090 GPU.\\
\subsection{Baselines and Evaluation Protocols}
\noindent\textbf{Baselines and Evaluation Dataset.} Few existing works employ neural network-free approaches for surface \textit{mesh-based} reconstruction in inverse rendering, primarily due to the difficulty of preserving surface topology as genus increases, introducing more tunnels and complex topological features. Many recent inverse rendering methods rely on alternative 3D representations. For example, Mehta et al.\cite{mehta2022level, mehta2023topologyderivative} introduced level-set-based representations for smooth surface reconstructions, while Munkberg et al.~\cite{Munkberg_2022Env} leveraged signed distance function (SDFs) within a neural framework to jointly optimize geometry, lighting, and materials. However, these methods are computationally expensive and not suitable for real-time applications. In contrast, surface \textit{mesh-based} representations offer a balanced trade-off between efficiency and geometric detail but continue to struggle with topology preservation. To evaluate our method’s ability to preserve surface topology in high-genus meshes, we compare against the state-of-the-art neural network-free approach by Nicolet et al.~\cite{Nicolet2021Large}, recently integrated into Mitsuba-3~\cite{Mitsuba3} and widely adopted as a strong baseline. Our evaluation includes five high-genus models, five genus-0 models from the Google Scanned Objects (GSO) dataset~\cite{downs2022google} and Stanford 3D Scanning Repository~\cite{stanford3d}, as well as high-genus shapes from Gu et al.~\cite{CCG08}, Crane et al.~\cite{Crane:2013:DGP}, and the Thingi10K dataset~\cite{Thingi10K}.

\noindent\textbf{Evaluation Metrics.}  
We evaluate reconstruction quality using two standard metrics: Chamfer Distance (CD) and Intersection over Union (IoU). Following established practice, we first apply Iterative Closest Point (ICP) alignment between the reconstructed mesh and the ground-truth surface prior to computing these metrics, which assess the geometric accuracy of the reconstructed surface mesh.

\input{fig/render_loss}

\input{tab/compare_combine}

\input{fig/curvature}

\subsection{Results}
\noindent\textbf{Multi-view Rendered Images.} Figures~\ref{fig:qualitative_highgenus} and~\ref{fig:qual_lowGenus} demonstrate that our method excels at reconstructing high-genus surfaces from multiple views, fully preserving topological features. In contrast, current state-of-the-art methods often fail to accurately capture these features, especially for high-genus surfaces. For low-genus surfaces, we show that our topologically informed approach enhances surface details and avoids visible seams, whereas existing methods frequently exhibit over-smoothing and cracks in regions of high curvature. These qualitative results highlight that relying solely on a reparameterized optimizer \cite{Nicolet2021Large}, without periodic coarsening and adaptive remeshing cycles, is insufficient for capturing key topological features. Tables~\ref{tab:quantitative_highgenus} present quantitative comparison results, further demonstrating that our method excels at reconstructing high-genus surfaces while preserving topological features and enhancing surface detail for low-genus surfaces, as measured by Chamfer Distance and Volume IoU. Additionally, the genus numbers in the tables show that the genus of the reconstructed surfaces matches that of the ground truth, ensuring topological consistency. Figure~\ref{fig:render_loss} tracks the \( l_1 \) rendering loss \( \Phi(R(\mathbf{x})) \) over optimization iterations, showing that our method generally converges in fewer iterations on average. For the Mario model, although the rendering loss plots converge to nearly zero, our method still achieves significantly better qualitative results. This demonstrates that a lower rendering loss does not necessarily correspond to a lower Chamfer Distance or a higher Volume IoU, while the reverse is generally true. The exception is the Planck model, where we set a limit of 1,500 iterations for evaluating low-genus surfaces. It requires more iteration to converge. Furthermore, Figure~\ref{fig:surface_detail} demonstrates that our method better preserves surface features, including curvature and geometric continuity, whereas the current state-of-the-art often produces noisy curvature and discontinuities.

%% file: fig/genus0.tex
\begin{figure*}[t]
    \centering
    \setlength{\resLen}{0.21\linewidth}
    \addtolength{\tabcolsep}{-6pt}
    \renewcommand{\arraystretch}{1.8} 
    \hskip -5pt
    \begin{tabular}{ccccccc}
        \raisebox{30pt}{\rotatebox[origin=c]{90}{Bunny}} & \hskip 5pt
        \begin{overpic}[height=0.65\resLen]{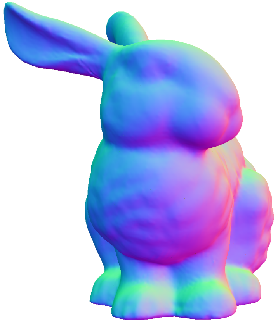}
            \put(0,0){
                \color{red}\linethickness{0.1pt}\polygon(17,75)(27,75)(27,87)(17,87)}
            \put(-15,25){
                \color{red}\linethickness{1pt}\frame{
                    \includegraphics[clip, trim=35 118 85 22, height=0.25\resLen]{img/genus0/bunny/nico_1.png}}}
        \end{overpic} & \hskip -15pt
        \begin{overpic}[height=0.65\resLen]{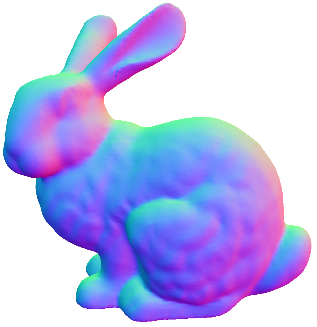}
            \put(0,0){
                \color{green}\linethickness{0.1pt}\polygon(25,70)(35,70)(35,85)(25,85)}
            \put(65,68){
                \color{green}\linethickness{1pt}\frame{
                    \includegraphics[clip, trim=50 112 90 28, height=0.25\resLen]{img/genus0/bunny/nico_2.png}}}
        \end{overpic} \hskip 5pt &
        \multicolumn{1}{|c}{ \hskip 5pt
        \begin{overpic}[height=0.65\resLen]{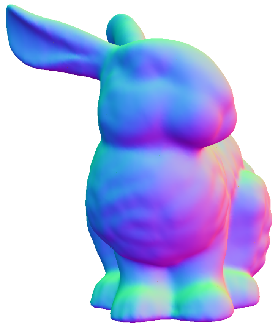}
            \put(0,0){
                \color{red}\linethickness{0.1pt}\polygon(17,75)(27,75)(27,87)(17,87)}
            \put(-15,25){
                \color{red}\linethickness{1pt}\frame{
                    \includegraphics[clip, trim=35 118 85 22, height=0.25\resLen]{img/genus0/bunny/ours_1.png}}}
        \end{overpic}} & \hskip -15pt
        \begin{overpic}[height=0.65\resLen]{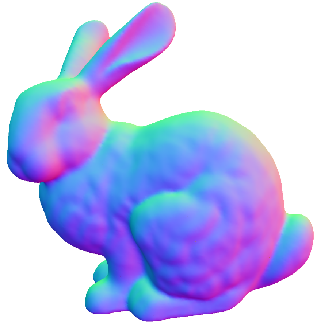}
            \put(0,0){
                \color{green}\linethickness{0.1pt}\polygon(25,70)(35,70)(35,85)(25,85)}
            \put(65,68){
                \color{green}\linethickness{1pt}\frame{
                    \includegraphics[clip, trim=50 112 90 28, height=0.25\resLen]{img/genus0/bunny/ours_2.png}}}
        \end{overpic} \hskip 5pt &
        \multicolumn{1}{|c}{ \hskip 5pt
        \begin{overpic}[height=0.65\resLen]{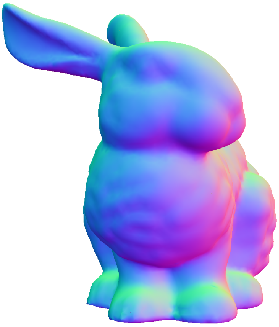}
            \put(0,0){
                \color{red}\linethickness{0.1pt}\polygon(17,75)(27,75)(27,87)(17,87)}
            \put(-15,25){
                \color{red}\linethickness{1pt}\frame{
                    \includegraphics[clip, trim=35 118 85 22, height=0.25\resLen]{img/genus0/bunny/gt_1.png}}}
        \end{overpic}} & \hskip -15pt
        \begin{overpic}[height=0.65\resLen]{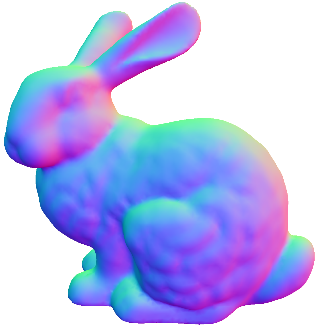}
            \put(0,0){
                \color{green}\linethickness{0.1pt}\polygon(25,70)(35,70)(35,85)(25,85)}
            \put(65,68){
                \color{green}\linethickness{1pt}\frame{
                    \includegraphics[clip, trim=52 112 88 28, height=0.25\resLen]{img/genus0/bunny/gt_2.png}}}
        \end{overpic} 
        \\
        \raisebox{40pt}{\rotatebox[origin=c]{90}{Nefertiti}} & \hskip 15pt
        \begin{overpic}[height=0.9\resLen]{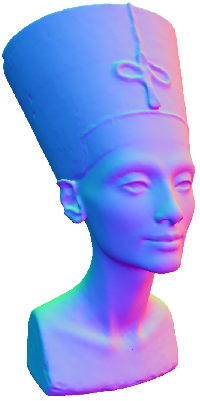}
            \put(0,0){
                \color{red}\linethickness{0.1pt}\polygon(10,43)(20,43)(20,56)(10,56)}
            \put(-20,25){
                \color{red}\linethickness{1pt}\frame{
                    \includegraphics[clip, trim=25 84 54 85, height=0.3\resLen]{img/genus0/nefertiti/nico_1.png}}}
        \end{overpic} & \hskip -15pt
        \begin{overpic}[height=0.9\resLen]{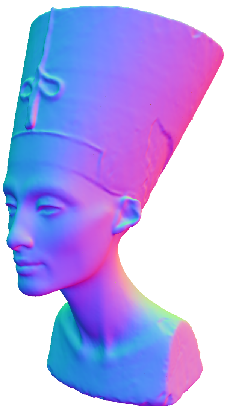}
            \put(0,0){
                \color{green}\linethickness{0.1pt}\polygon(25,42)(33,42)(33,55)(25,55)}
            \put(40,25){
                \color{green}\linethickness{1pt}\frame{
                    \includegraphics[clip, trim=55 83 40 91, height=0.3\resLen]{img/genus0/nefertiti/nico_2.png}}}
        \end{overpic} \hskip 5pt & 
        \multicolumn{1}{|c}{\hskip 15pt
        \begin{overpic}[height=0.9\resLen]{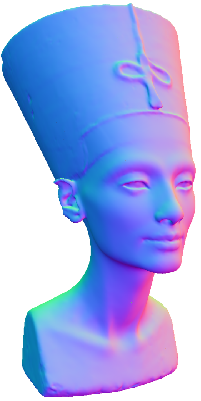}
            \put(0,0){
                \color{red}\linethickness{0.1pt}\polygon(10,43)(20,43)(20,56)(10,56)}
            \put(-20,25){
                \color{red}\linethickness{1pt}\frame{
                    \includegraphics[clip, trim=25 84 54 85, height=0.3\resLen]{img/genus0/nefertiti/ours_1.png}}}
        \end{overpic}} & \hskip -15pt
        \begin{overpic}[height=0.9\resLen]{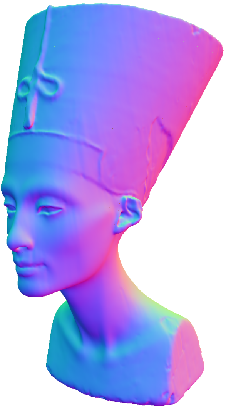}
            \put(0,0){
                \color{green}\linethickness{0.1pt}\polygon(25,42)(33,42)(33,55)(25,55)}
            \put(40,25){
                \color{green}\linethickness{1pt}\frame{
                    \includegraphics[clip, trim=55 83 40 91, height=0.3\resLen]{img/genus0/nefertiti/ours_2.png}}}
        \end{overpic} \hskip 5pt & 
        \multicolumn{1}{|c}{\hskip 15pt
        \begin{overpic}[height=0.9\resLen]{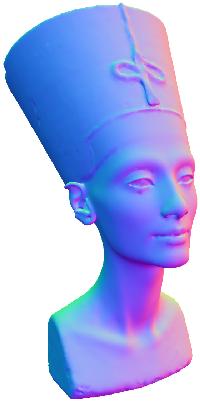}
            \put(0,0){
                \color{red}\linethickness{0.1pt}\polygon(12,43)(22,43)(22,56)(12,56)}
            \put(-20,25){
                \color{red}\linethickness{1pt}\frame{
                    \includegraphics[clip, trim=30 84 50 85, height=0.3\resLen]{img/genus0/nefertiti/gt_1.png}}}
        \end{overpic}} & \hskip -15pt
        \begin{overpic}[height=0.9\resLen]{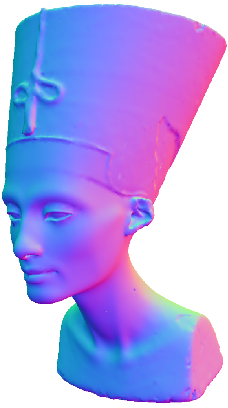}
            \put(0,0){
                \color{green}\linethickness{0.1pt}\polygon(28,42)(36,42)(36,55)(28,55)}
            \put(40,25){
                \color{green}\linethickness{1pt}\frame{
                    \includegraphics[clip, trim=60 83 35 91, height=0.3\resLen]{img/genus0/nefertiti/gt_2.png}}}
        \end{overpic} 
        \\
        \raisebox{40pt}{\rotatebox[origin=c]{90}{Planck}} &
        \begin{overpic}[height=0.8\resLen]{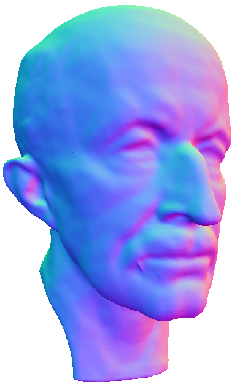}
            \put(0,0){
                \color{green}\linethickness{0.1pt}\polygon(28,24)(38,24)(38,37)(28,37)}
            \put(60,0){
                \color{green}\linethickness{1pt}\frame{
                    \includegraphics[clip, trim=60 46 38 119, height=0.3\resLen]{img/genus0/planck/nico_2.png}}}
        \end{overpic} &
        \begin{overpic}[height=0.8\resLen]{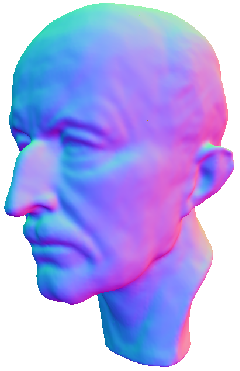}
        \end{overpic} \hskip 5pt &
        \multicolumn{1}{|c}{
        \begin{overpic}[height=0.8\resLen]{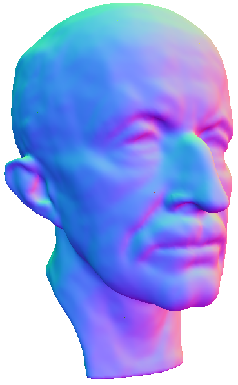}
            \put(0,0){
                \color{green}\linethickness{0.1pt}\polygon(28,24)(38,24)(38,37)(28,37)}
            \put(60,0){
                \color{green}\linethickness{1pt}\frame{
                    \includegraphics[clip, trim=60 46 38 119, height=0.3\resLen]{img/genus0/planck/ours_2.png}}}
        \end{overpic}} &
        \begin{overpic}[height=0.8\resLen]{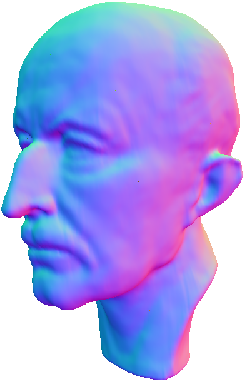}
        \end{overpic} \hskip 5pt &
        \multicolumn{1}{|c}{
        \begin{overpic}[height=0.8\resLen]{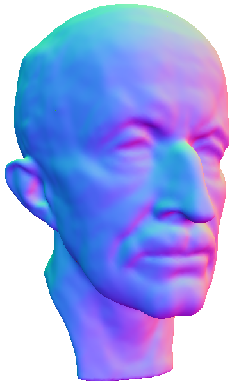}
            \put(0,0){
                \color{green}\linethickness{0.1pt}\polygon(28,24)(38,24)(38,37)(28,37)}
            \put(60,0){
                \color{green}\linethickness{1pt}\frame{
                    \includegraphics[clip, trim=60 46 38 119, height=0.3\resLen]{img/genus0/planck/gt_2.png}}}
        \end{overpic}} &
        \begin{overpic}[height=0.8\resLen]{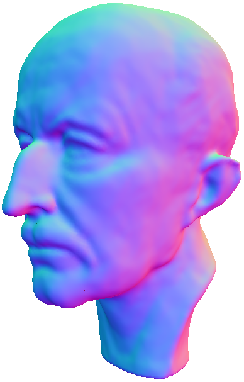}
        \end{overpic} 
        \\
        \raisebox{40pt}{\rotatebox[origin=c]{90}{Mario}} &
        \begin{overpic}[height=0.95\resLen]{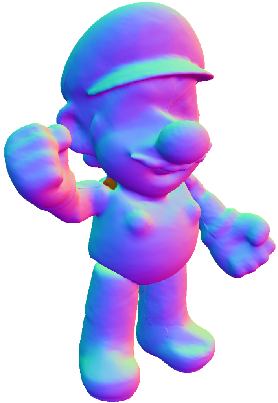}
            \put(0,0){
                \color{red}\linethickness{0.1pt}\polygon(20,50)(30,50)(30,62)(20,62)}
            \put(-10,5){
                \color{red}\linethickness{1pt}\frame{
                    \includegraphics[clip, trim=45 95 72 75, height=0.3\resLen]{img/genus0/mario/nico_1.png}}}
        \end{overpic} &
        \begin{overpic}[height=0.95\resLen]{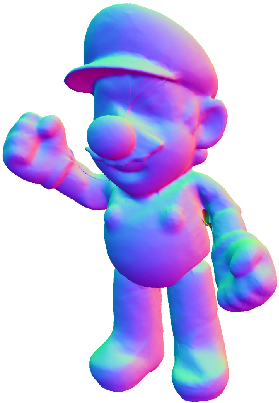}
            \put(0,0){
                \color{green}\linethickness{0.1pt}\polygon(45,40)(55,40)(55,52)(45,52)}
            \put(-10,5){
                \color{green}\linethickness{1pt}\frame{
                    \includegraphics[clip, trim=90 75 25 95, height=0.3\resLen]{img/genus0/mario/nico_2.png}}}
        \end{overpic} \hskip 5pt &
        \multicolumn{1}{|c}{
        \begin{overpic}[height=0.95\resLen]{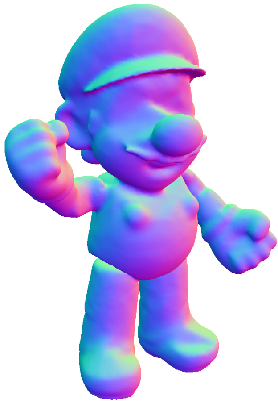}
            \put(0,0){
                \color{red}\linethickness{0.1pt}\polygon(20,50)(30,50)(30,62)(20,62)}
            \put(-10,5){
                \color{red}\linethickness{1pt}\frame{
                    \includegraphics[clip, trim=45 95 72 75, height=0.3\resLen]{img/genus0/mario/ours_1.png}}}
        \end{overpic}} &
        \begin{overpic}[height=0.95\resLen]{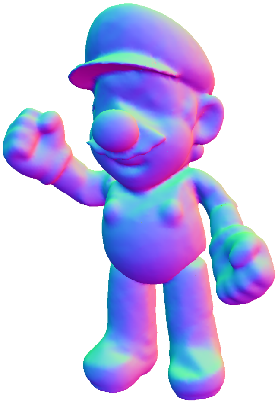}
            \put(0,0){
                \color{green}\linethickness{0.1pt}\polygon(45,40)(55,40)(55,52)(45,52)}
            \put(-10,5){
                \color{green}\linethickness{1pt}\frame{
                    \includegraphics[clip, trim=90 75 25 95, height=0.3\resLen]{img/genus0/mario/ours_2.png}}}
        \end{overpic} \hskip 5pt &
        \multicolumn{1}{|c}{
        \begin{overpic}[height=0.95\resLen]{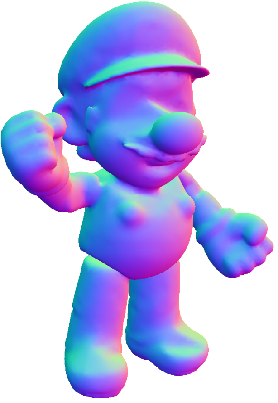}
            \put(0,0){
                \color{red}\linethickness{0.1pt}\polygon(20,50)(30,50)(30,62)(20,62)}
            \put(-12,5){
                \color{red}\linethickness{1pt}\frame{
                    \includegraphics[clip, trim=45 95 72 75, height=0.3\resLen]{img/genus0/mario/gt_1.png}}}
        \end{overpic}} &
        \begin{overpic}[height=0.95\resLen]{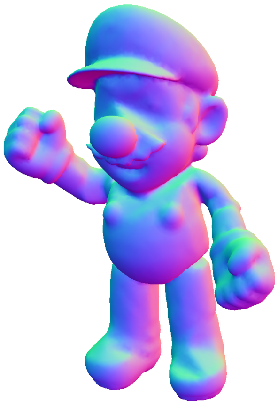}
            \put(0,0){
                \color{green}\linethickness{0.1pt}\polygon(45,40)(55,40)(55,52)(45,52)}
            \put(-10,5){
                \color{green}\linethickness{1pt}\frame{
                    \includegraphics[clip, trim=90 75 25 95, height=0.3\resLen]{img/genus0/mario/gt_2.png}}}
        \end{overpic} 
        \\
        \raisebox{40pt}{\rotatebox[origin=c]{90}{Armadillo}} &
        \begin{overpic}[height=\resLen]{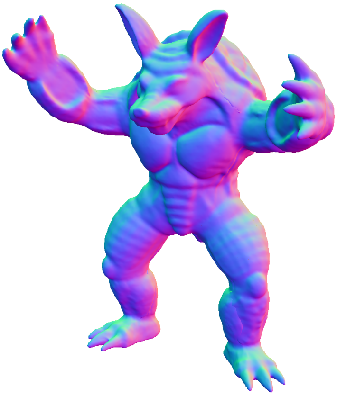}
            \put(0,0){
                \color{red}\linethickness{0.1pt}\polygon(50,70)(60,70)(60,82)(50,82)}
            \put(0,30){
                \color{red}\linethickness{1pt}\frame{
                    \includegraphics[clip, trim=100 130 42 35, height=0.3\resLen]{img/genus0/armadillo/nico_1.png}}}
        \end{overpic} &
        \begin{overpic}[height=\resLen]{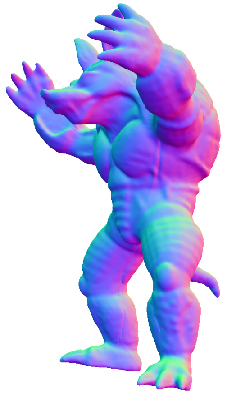}
            \put(0,0){
                \color{green}\linethickness{0.1pt}\polygon(23,13)(33,13)(33,25)(23,25)}
            \put(-10,25){
                \color{green}\linethickness{1pt}\frame{
                    \includegraphics[clip, trim=50 25 42 145, height=0.3\resLen]{img/genus0/armadillo/nico_2.png}}}
        \end{overpic} \hskip 5pt &
        \multicolumn{1}{|c}{
        \begin{overpic}[height=\resLen]{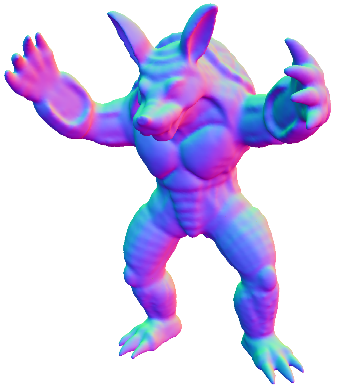}
            \put(0,0){
                \color{red}\linethickness{0.1pt}\polygon(50,70)(60,70)(60,82)(50,82)}
            \put(0,30){
                \color{red}\linethickness{1pt}\frame{
                    \includegraphics[clip, trim=100 130 46 35, height=0.3\resLen]{img/genus0/armadillo/ours_1.png}}}
        \end{overpic}} &
        \begin{overpic}[height=\resLen]{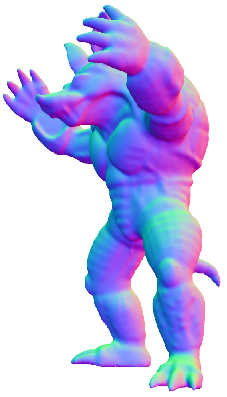}
            \put(0,0){
                \color{green}\linethickness{0.1pt}\polygon(25,13)(35,13)(35,25)(25,25)}
            \put(-10,25){
                \color{green}\linethickness{1pt}\frame{
                    \includegraphics[clip, trim=55 25 38 145, height=0.3\resLen]{img/genus0/armadillo/ours_2.png}}}
        \end{overpic} \hskip 5pt &
        \multicolumn{1}{|c}{
        \begin{overpic}[height=\resLen]{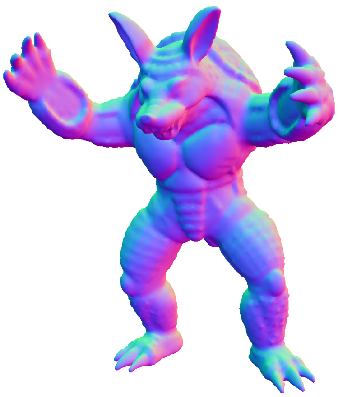}
            \put(0,0){
                \color{red}\linethickness{0.1pt}\polygon(50,70)(60,70)(60,82)(50,82)}
            \put(0,30){
                \color{red}\linethickness{1pt}\frame{
                    \includegraphics[clip, trim=100 130 42 35, height=0.3\resLen]{img/genus0/armadillo/gt_1.png}}}
        \end{overpic}} &
        \begin{overpic}[height=\resLen]{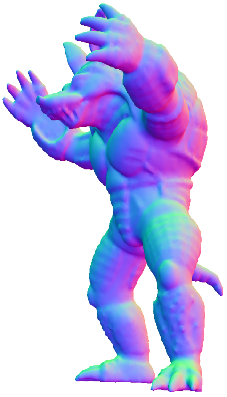}
            \put(0,0){
                \color{green}\linethickness{0.1pt}\polygon(25,13)(35,13)(35,25)(25,25)}
            \put(-10,25){
                \color{green}\linethickness{1pt}\frame{
                    \includegraphics[clip, trim=55 25 38 145, height=0.3\resLen]{img/genus0/armadillo/gt_2.png}}}
        \end{overpic} 
        \\
        & \multicolumn{2}{c}{Nicolet et al.\cite{Nicolet2021Large}} & \multicolumn{2}{c}{Ours} & \multicolumn{2}{c}{GT} 
    \end{tabular}
    \caption{Qualitative Results of Multi-View Reconstruction for Genus 0 Surfaces with Rendered Views and Normal Maps Using a Sphere (\textbf{Genus 0}) as Topological Primitive. Please see \textbf{Appendix \textcolor{red}{A.1}} for the complete set of multi-view genus-0 surface qualitative results.}
    \label{fig:qual_lowGenus}
\end{figure*}

%% file: fig/render_loss.tex
\begin{figure*}[t] 
    \centering
    \setlength{\resLen}{0.2\linewidth}
    \setlength{\tabcolsep}{0pt}
    \begin{tabular}{ccccc} 
        \includegraphics[width=\resLen]{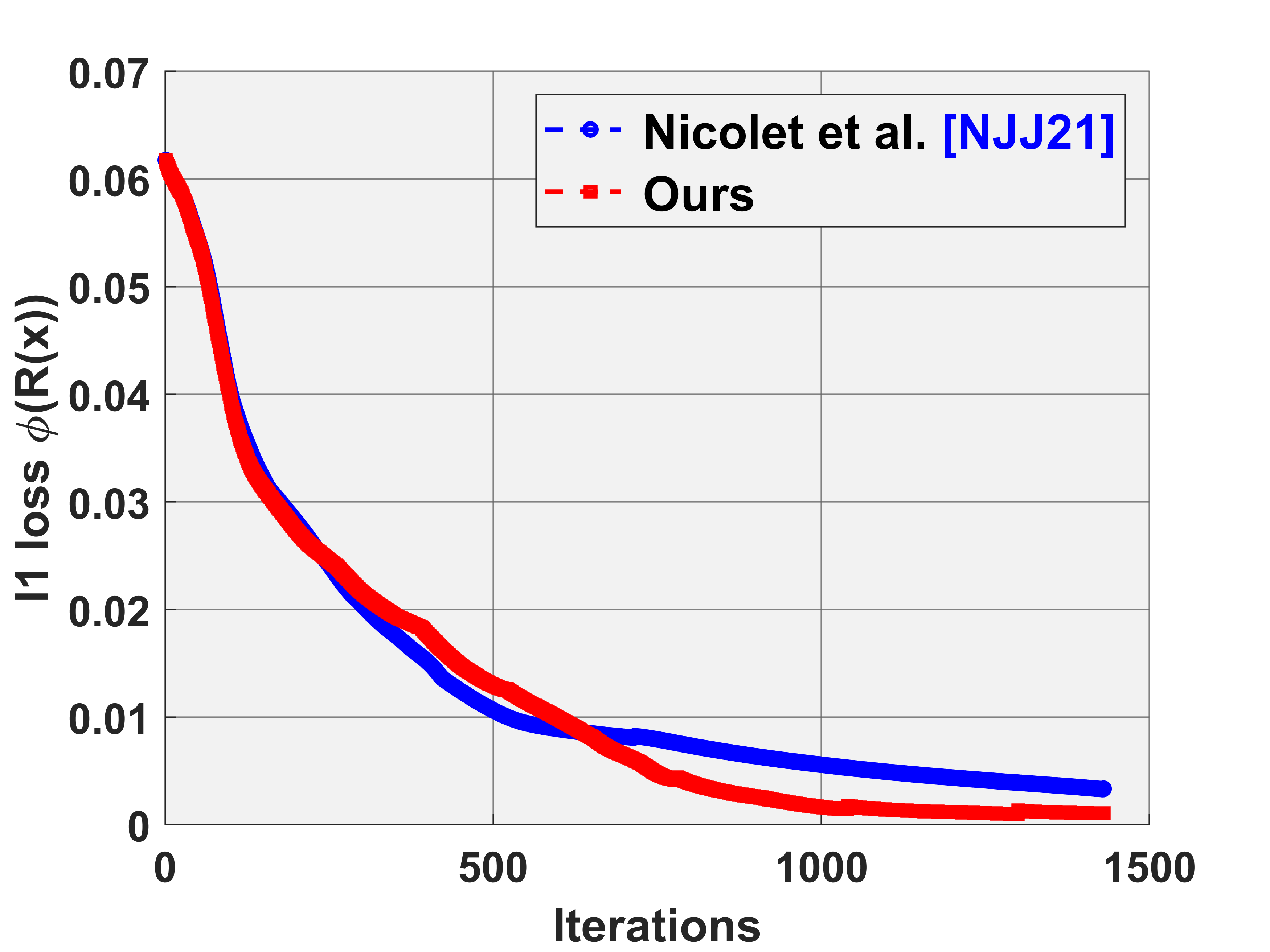} & 
        \includegraphics[width=\resLen]{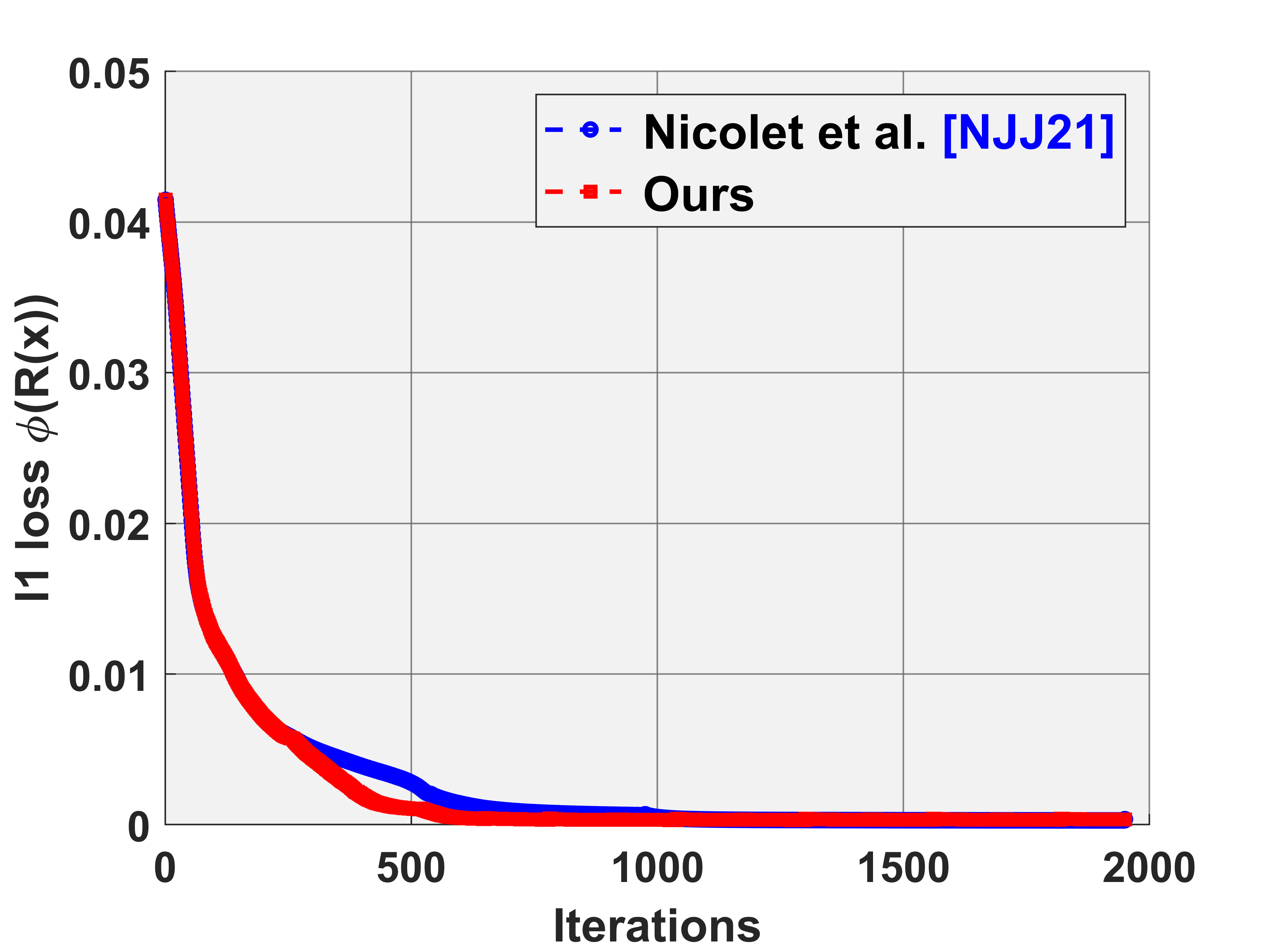} &
        \includegraphics[width=\resLen]{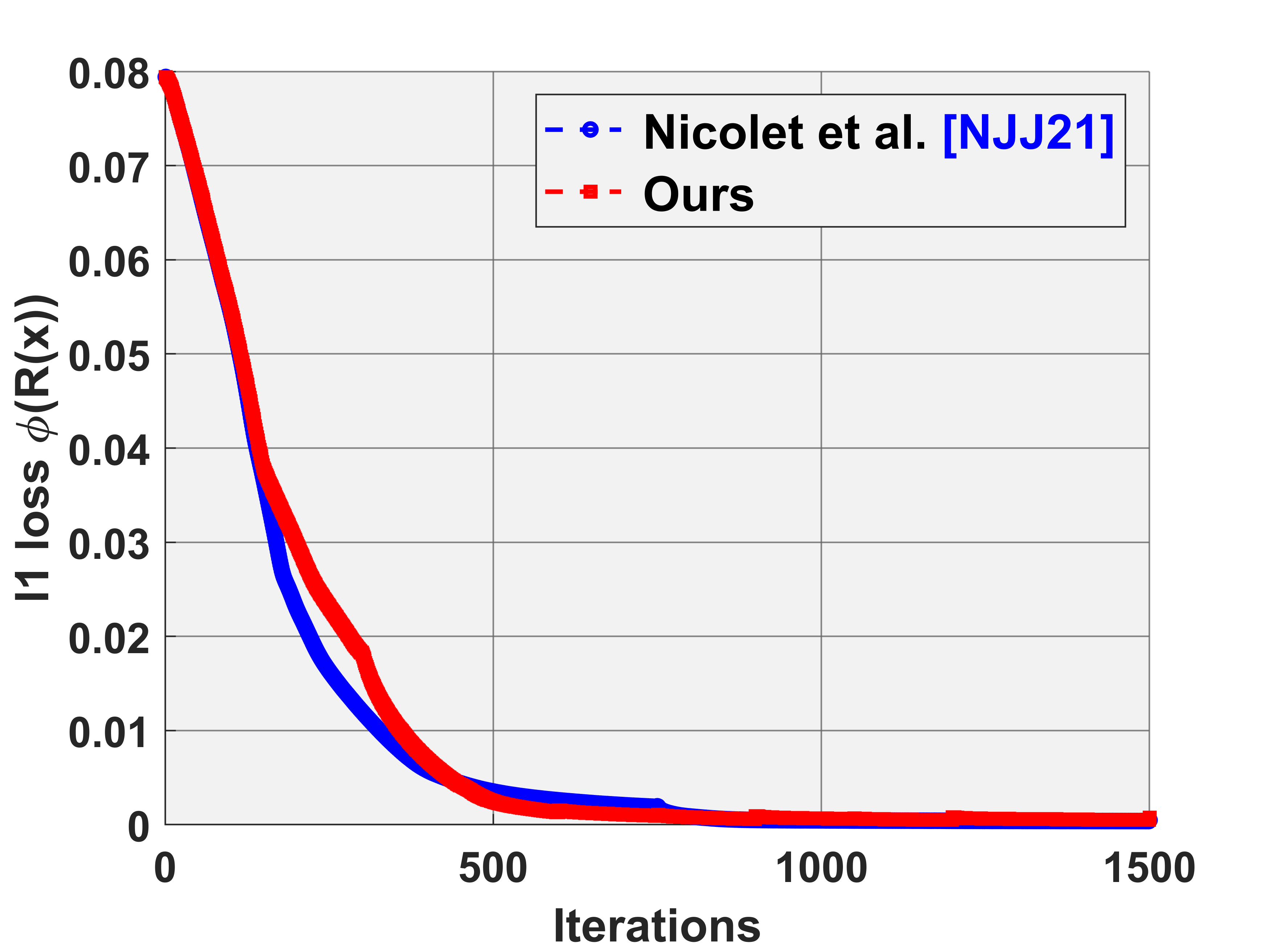} &
        \includegraphics[width=\resLen]{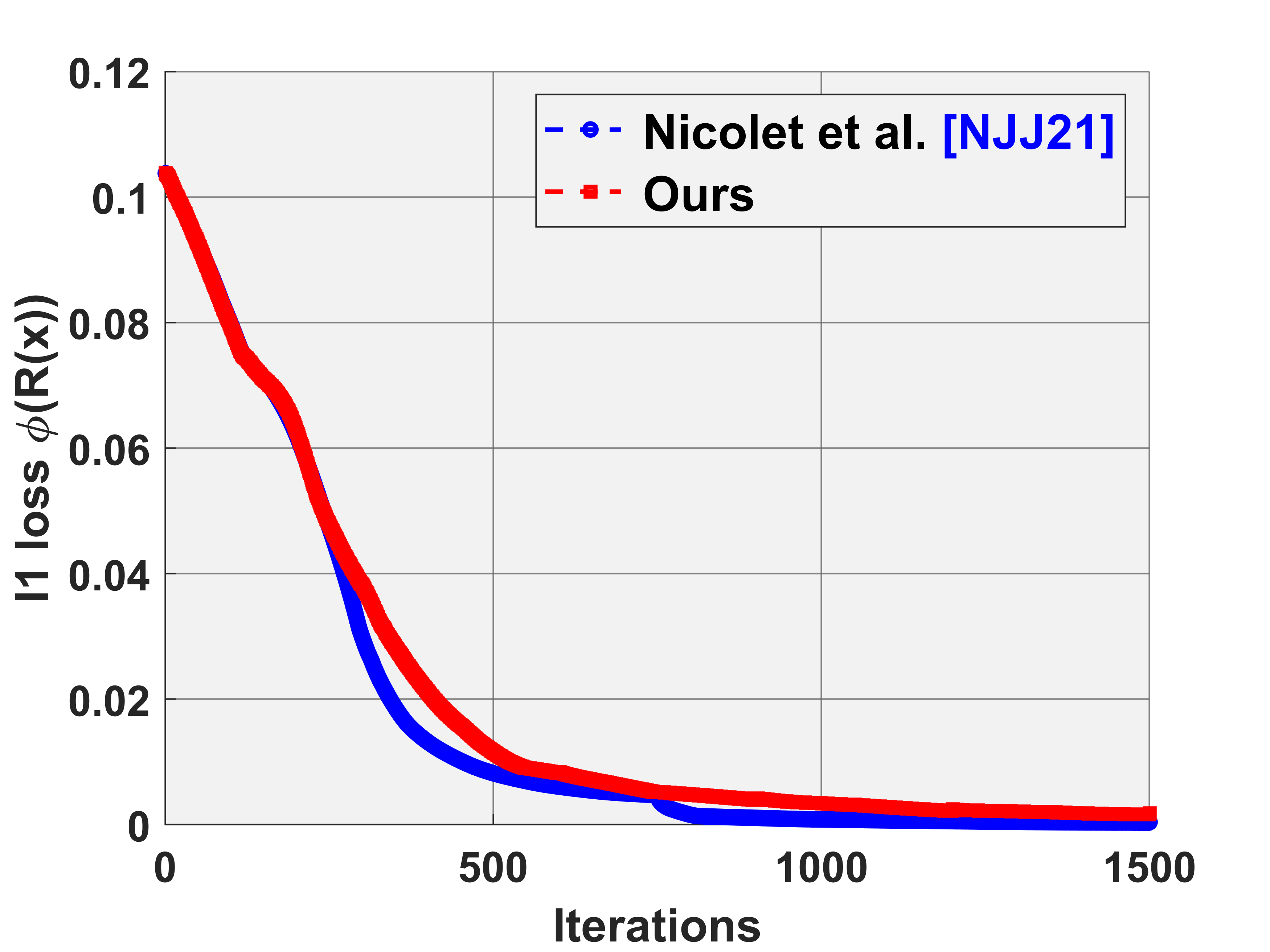} &
        \includegraphics[width=\resLen]{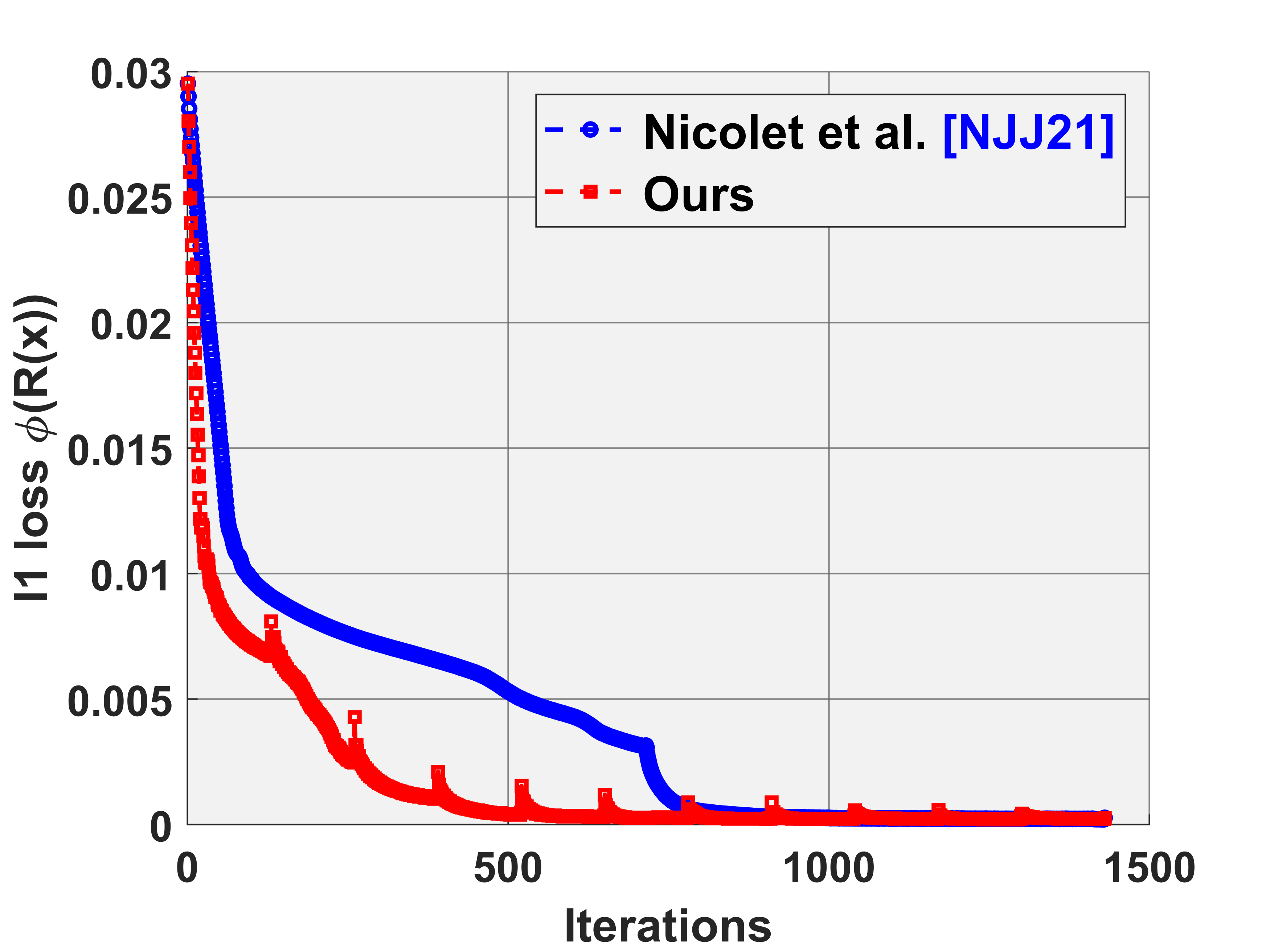} \\
        a1) Armadillo & a2) Bunny & a3) Nefertiti & a4) Planck & a5) Mario \\
        \includegraphics[width=\resLen]{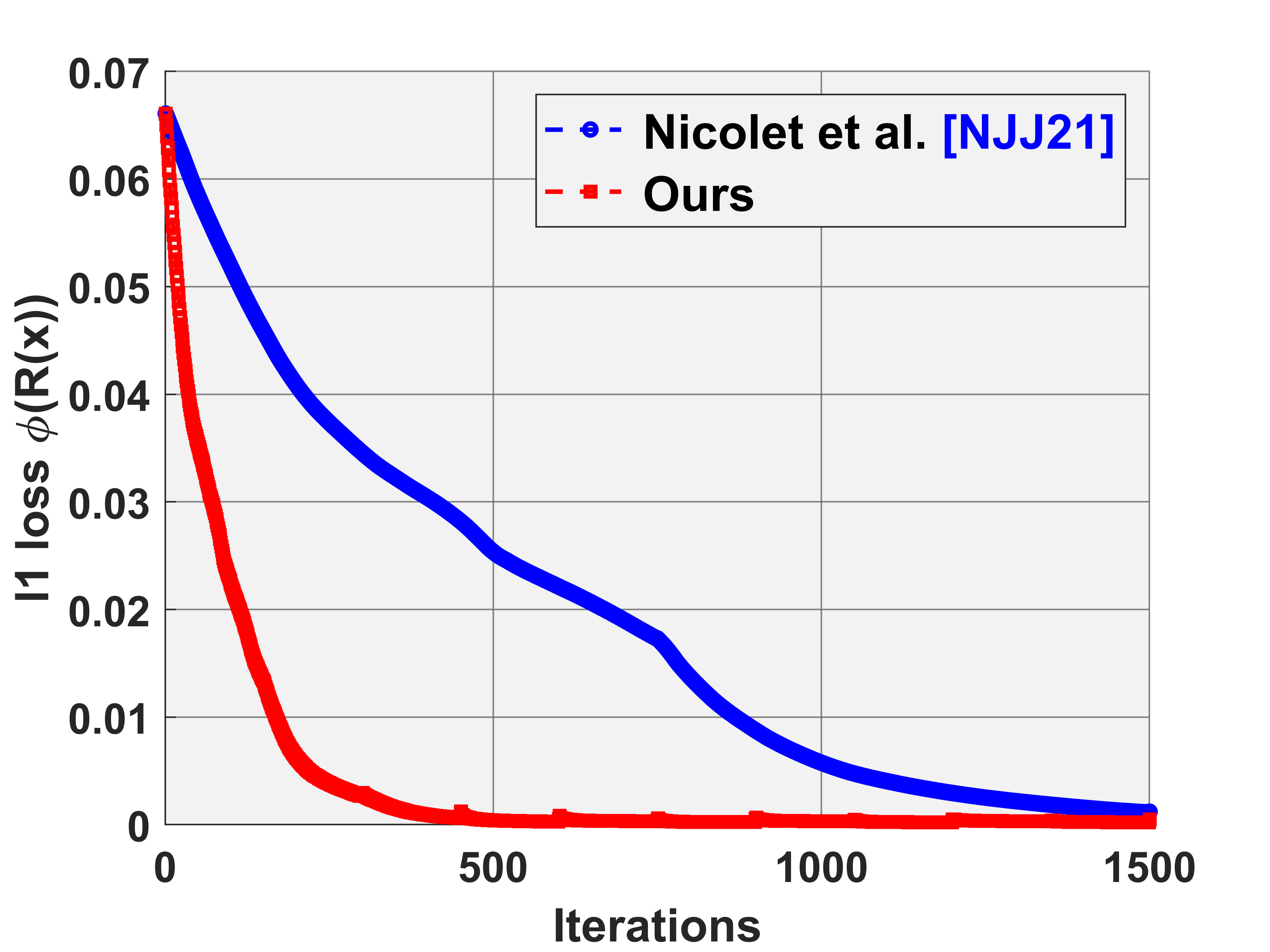} & 
        \includegraphics[width=\resLen]{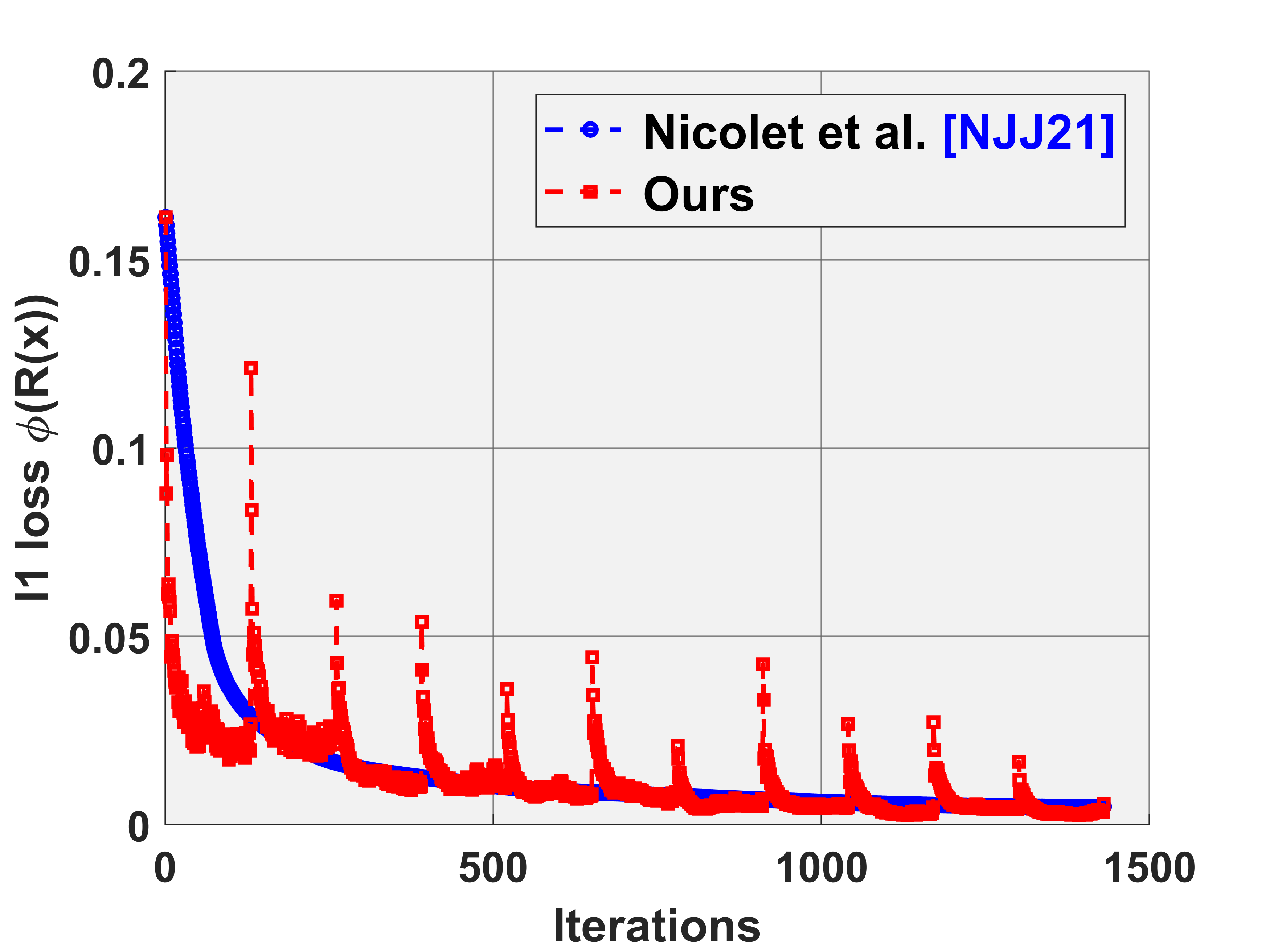} &
        \includegraphics[width=\resLen]{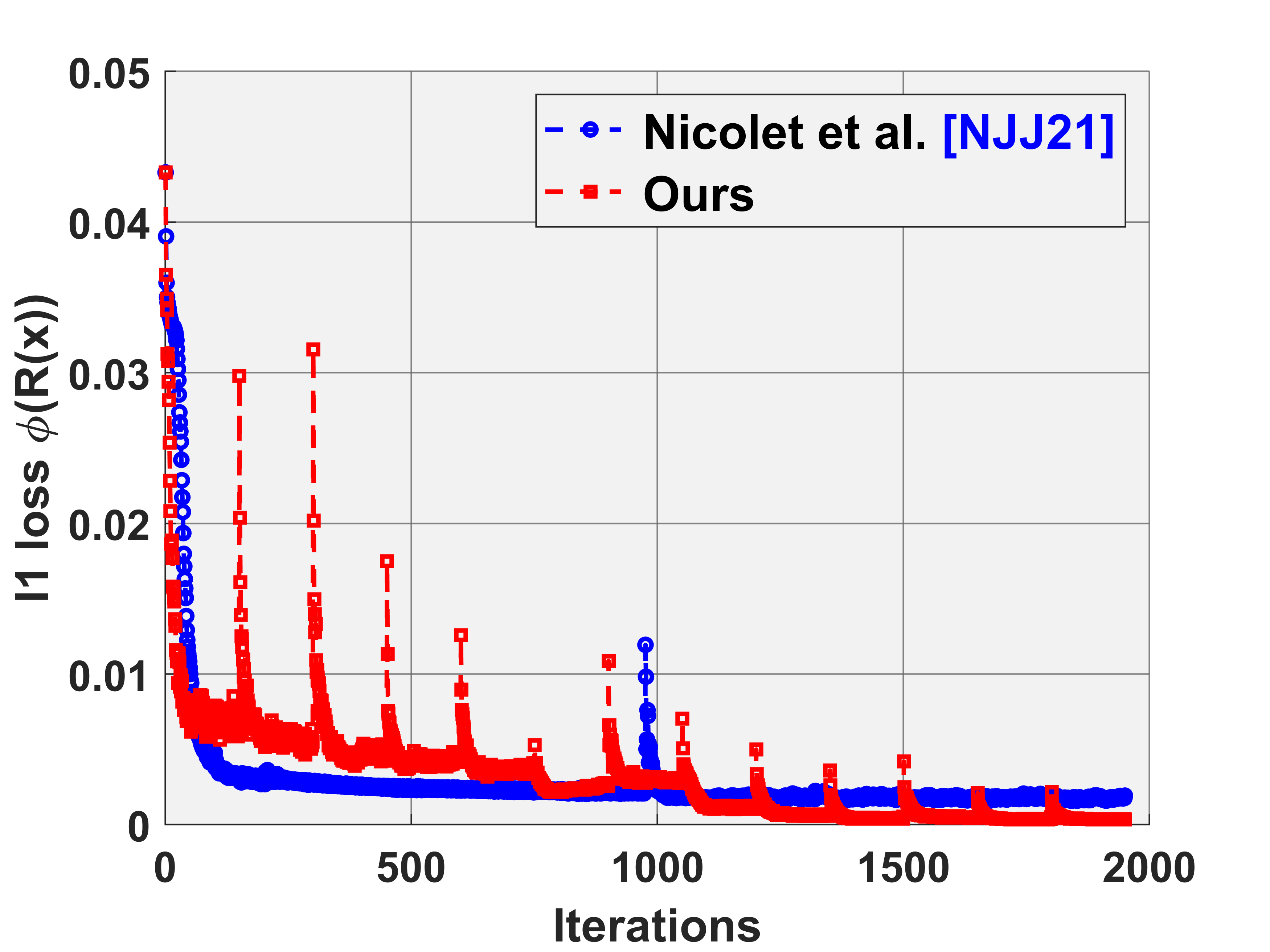} &
        \includegraphics[width=\resLen]{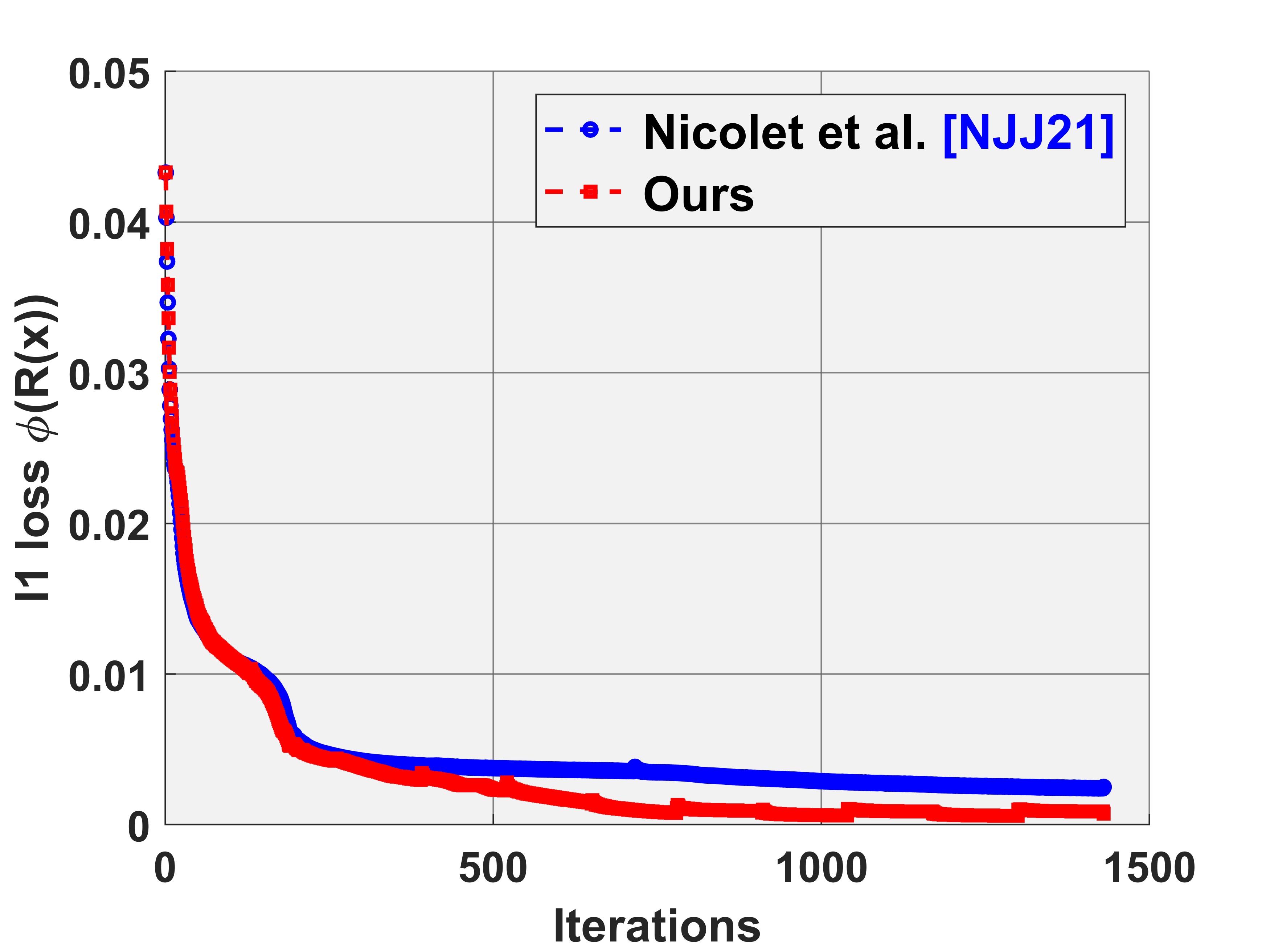} & 
        \includegraphics[width=\resLen]{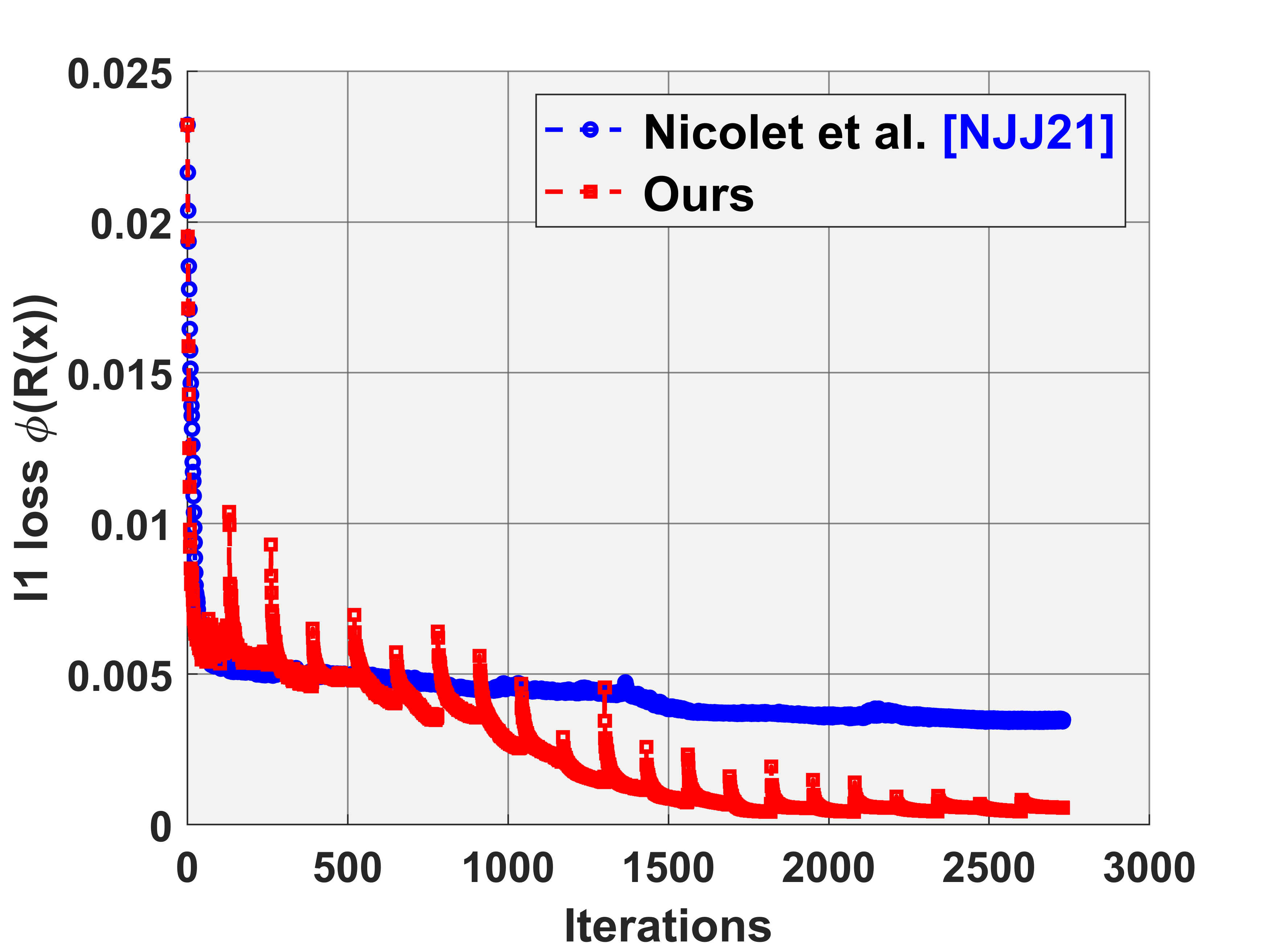} \\
        b1) Kitten & b2) Amphora & b3) Pretzel & b4) Birthday & b5) Sorter
    \end{tabular}
    \caption{Rendering loss for low-genus surfaces (\textbf{Top Row}: Armadillo, Bunny, Nefertiti, Planck, Mario) and high-genus surfaces (\textbf{Bottom Row}: Kitten, Amphora, Pretzel, Birthday, Sorter) are shown, where \textbf{\textcolor{red}{red}} indicates our results and \textcolor{blue}{blue} represents Nicolet et al.~\cite{Nicolet2021Large}.}
    \label{fig:render_loss}
\end{figure*}

%% file: tab/compare_combine.tex
\begin{table}[t]
    \centering
    \setlength{\tabcolsep}{4.0pt} 
    \renewcommand{\arraystretch}{1.025} 
    \small 
    \caption{Quantitative comparison with Nicolet et al. \cite{Nicolet2021Large} for multi-view reconstructed low-genus and high-genus surfaces: Lower Chamfer Distance values means better result while higher Volume IoU means better result. \textbf{Please see \textbf{Appendix \textcolor{red}{D}} for full table.}}
    
    \begin{tabular}{|p{2pt}|c|cc|cc|c|c|}
        \hline
        & \multirow{2}{*}{\textbf{Model}} & \multicolumn{2}{c|}{CD \( \downarrow \)} & \multicolumn{2}{c|}{IoU \( \uparrow \)} & \( \chi(S) \) & \( g \) \\
        \cline{3-8}
        & & \cite{Nicolet2021Large} & \textbf{Ours} & \cite{Nicolet2021Large} & \textbf{Ours} & \textbf{Ours} & \textbf{Ours} \\
        \hline
        \multirow{5}{*}{\hskip -2pt \rotatebox[origin=c]{90}{\textbf{Genus=0}}} & 
        Armadillo & 0.0018 & \textbf{0.0015} & 0.8968 & \textbf{0.9283} & 2 & 0 \\
        & Bunny & 0.0021 & \textbf{0.0020} & 0.8263 & \textbf{0.8296} & 2 & 0 \\
        & Nefertiti & 0.0018 & \textbf{0.0017} & 0.8768 & \textbf{0.9158} & 2 & 0 \\
        & Planck & 0.0019 & \textbf{0.0018} & \textbf{0.9369} & 0.9261 & 2 & 0 \\
        & Mario & \textbf{0.0024} & \textbf{0.0024} & 0.8683 & \textbf{0.9003} & 2 & 0 \\
        \hline
        \multirow{5}{*}{\hskip -2pt \rotatebox[origin=c]{90}{\textbf{Higher-Genus}}} & 
        Kitten & 0.0039 & \textbf{0.0025} & 0.6298 & \textbf{0.7126} & 0 & 1 \\
        & Amphora & 0.0054 & \textbf{0.0033} & 0.4581 & \textbf{0.7924} & 2 & 2 \\
        & Pretzel & 0.0040 & \textbf{0.0025} & 0.6518 & \textbf{0.8639} & 4 & 3 \\
        & Birthday & 0.0020 & \textbf{0.0006} & 0.4914 & \textbf{0.8849} & 6 & 4 \\
        & Sorter & 0.0672 & \textbf{0.0040} & 0.2901 & \textbf{0.7504} & 8 & 5 \\
        \hline
    \end{tabular}
    \label{tab:quantitative_highgenus}
\end{table}

%% file: fig/curvature.tex
\begin{figure}[t]
    \centering
    \setlength{\resLen}{0.2\linewidth}
    \addtolength{\tabcolsep}{4pt}
    \hskip -12pt
    \begin{tabular}{c@{\hskip 4pt}ccc} 
        \raisebox{40pt}{\thead{Mean\\Curvature\\H}} & 
        \begin{overpic}[width=\resLen]{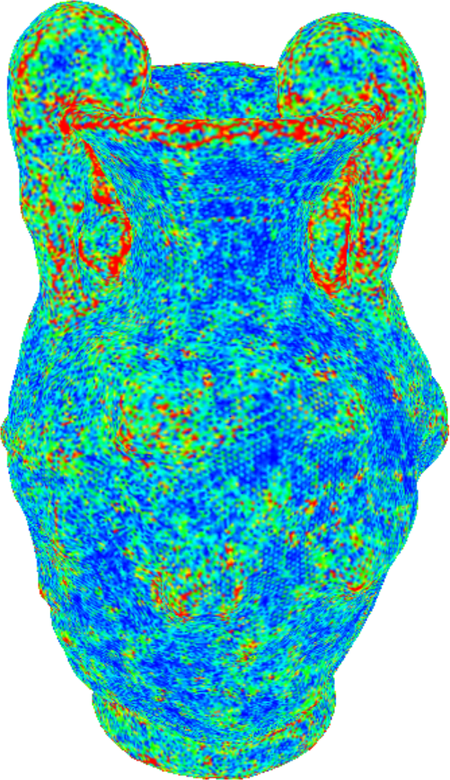}
            \put(0,0){
                \color{blue}\linethickness{0.1pt}
                    \polygon(15,15)(30,15)(30,30)(15,30)}
            \put(40,0){
                \color{blue}\linethickness{1pt}\frame{
                    \includegraphics[clip, trim=140 120 200 550, width=0.5\resLen]{img/curvature/h_nico.png}}} 
        \end{overpic} &
        \begin{overpic}[width=\resLen]{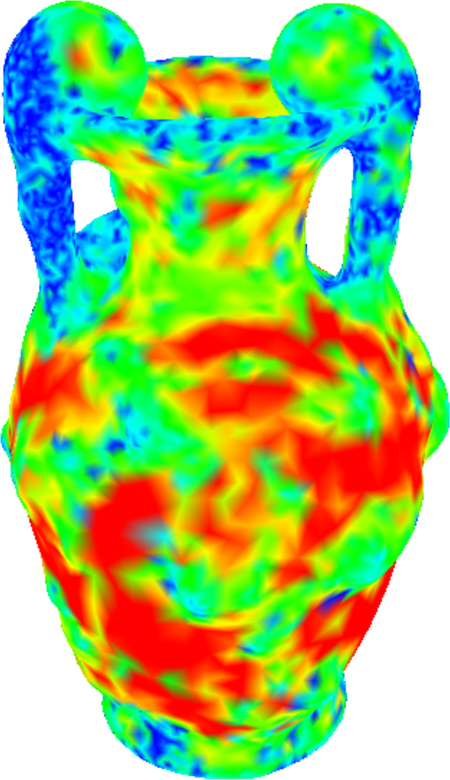}
            \put(0,0){
                \color{blue}\linethickness{0.1pt}
                    \polygon(15,15)(30,15)(30,30)(15,30)}
            \put(38,0){
                \color{blue}\linethickness{1pt}\frame{
                    \includegraphics[clip, trim=140 120 200 550, width=0.5\resLen]{img/curvature/h_ours.png}}} 
        \end{overpic} &
        \begin{overpic}[width=\resLen]{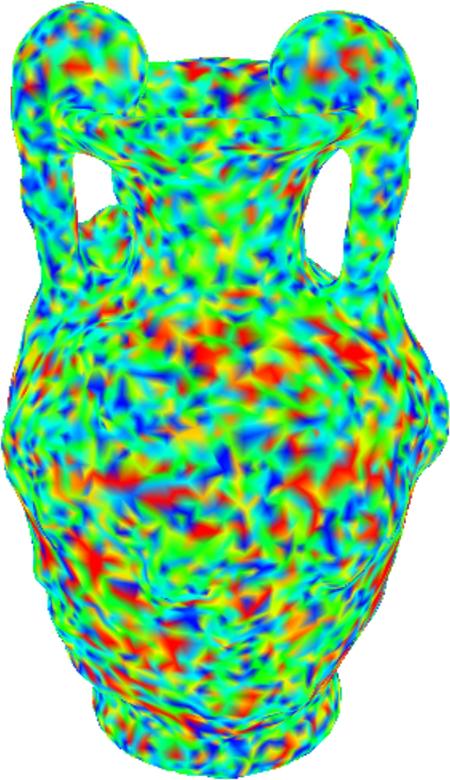}
            \put(0,0){
                \color{blue}\linethickness{0.1pt}
                    \polygon(15,15)(30,15)(30,30)(15,30)}
            \put(40,0){
                \color{blue}\linethickness{1pt}\frame{
                    \includegraphics[clip, trim=140 120 200 550, width=0.5\resLen]{img/curvature/h_gt.png}}} 
        \end{overpic}
        \\
        \raisebox{40pt}{\thead{Gaussian\\Curvature\\K}} & 
        \begin{overpic}[width=\resLen]{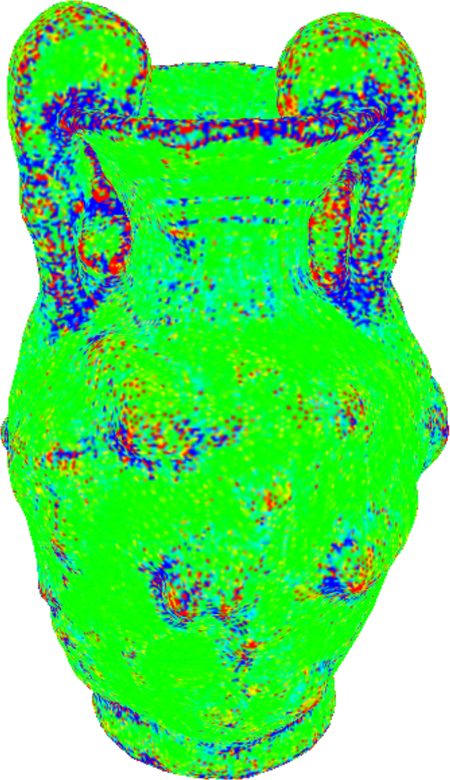}
            \put(0,0){
                \color{blue}\linethickness{0.1pt}
                    \polygon(32,84)(47,84)(47,99)(32,99)}
            \put(40,0){
                \color{blue}\linethickness{1pt}\frame{
                    \includegraphics[clip, trim=270 650 70 10, width=0.5\resLen]{img/curvature/g_nico.png}}} 
        \end{overpic} &
        \begin{overpic}[width=\resLen]{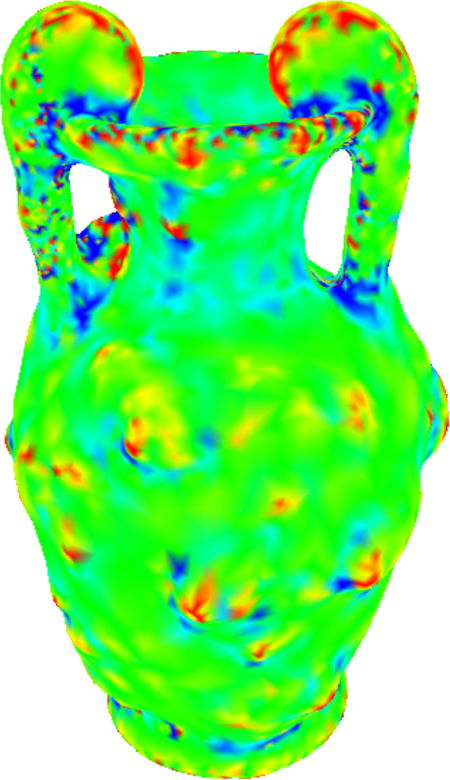}
            \put(0,0){
                \color{blue}\linethickness{0.1pt}
                    \polygon(32,84)(47,84)(47,99)(32,99)}
            \put(38,0){
                \color{blue}\linethickness{1pt}\frame{
                    \includegraphics[clip, trim=270 650 70 10, width=0.5\resLen]{img/curvature/g_ours.png}}} 
        \end{overpic} &
        \begin{overpic}[width=\resLen]{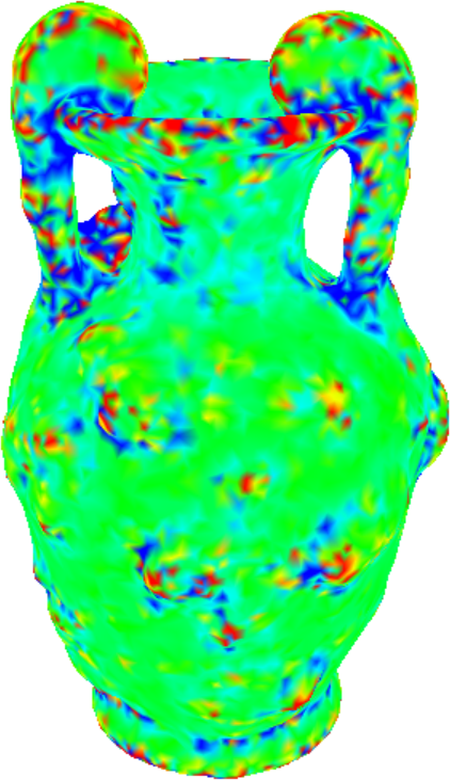}
            \put(0,0){
                \color{blue}\linethickness{0.1pt}
                    \polygon(32,84)(47,84)(47,99)(32,99)}
            \put(40,0){
                \color{blue}\linethickness{1pt}\frame{
                    \includegraphics[clip, trim=270 650 70 10, width=0.5\resLen]{img/curvature/g_gt.png}}} 
        \end{overpic}
        \\
        \raisebox{40pt}{\thead{Surface\\Continuity}} & 
        \begin{overpic}[width=\resLen]{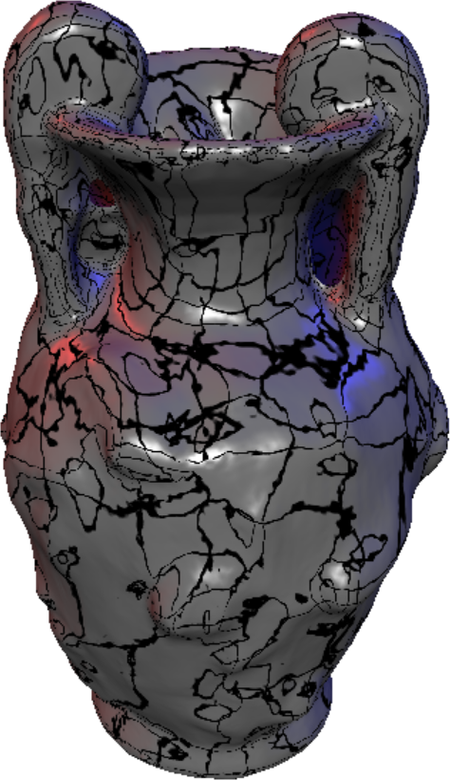}
            \put(0,0){
                \color{blue}\linethickness{0.1pt}
                    \polygon(15,15)(30,15)(30,30)(15,30)}
            \put(40,0){
                \color{blue}\linethickness{1pt}\frame{
                    \includegraphics[clip, trim=140 120 200 550, width=0.5\resLen]{img/curvature/c_nico.png}}} 
        \end{overpic} &
        \begin{overpic}[width=\resLen]{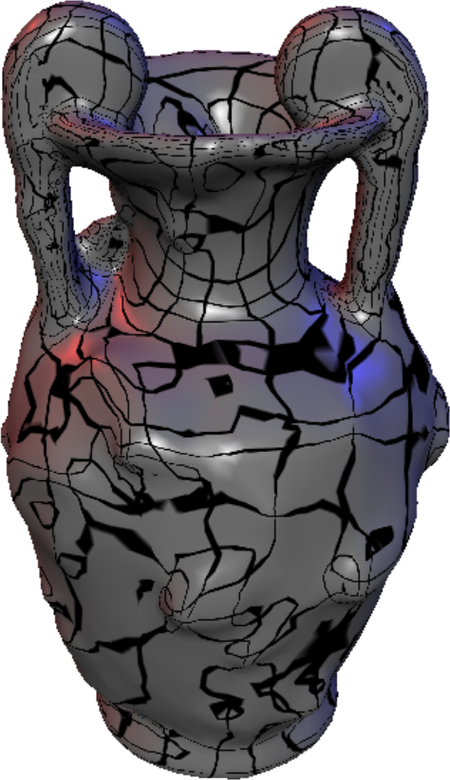}
            \put(0,0){
                \color{blue}\linethickness{0.1pt}
                    \polygon(15,15)(30,15)(30,30)(15,30)}
            \put(38,0){
                \color{blue}\linethickness{1pt}\frame{
                    \includegraphics[clip, trim=140 120 200 550, width=0.5\resLen]{img/curvature/c_ours.png}}} 
        \end{overpic} &
        \begin{overpic}[width=\resLen]{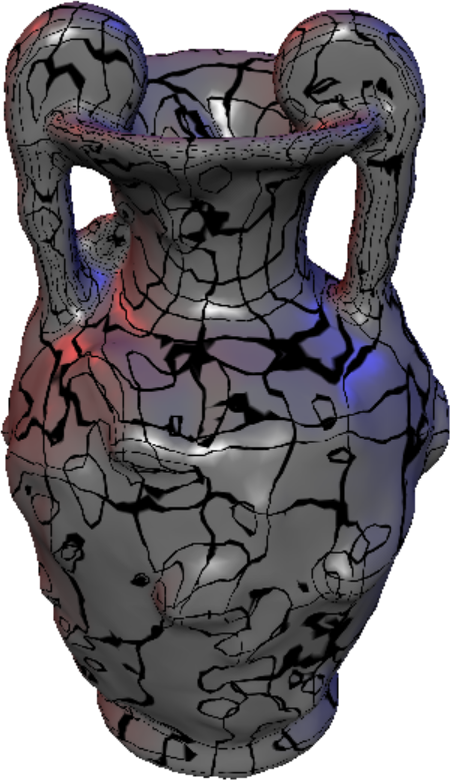}
            \put(0,0){
                \color{blue}\linethickness{0.1pt}
                    \polygon(15,15)(30,15)(30,30)(15,30)}
            \put(40,0){
                \color{blue}\linethickness{1pt}\frame{
                    \includegraphics[clip, trim=140 120 200 550, width=0.5\resLen]{img/curvature/c_gt.png}}} 
        \end{overpic}
        \\
        & \thead{Nicolet et al.\\\cite{Nicolet2021Large}} & Ours & GT
    \end{tabular}\vspace{-5pt}
    \caption{Qualitative comparison with Nicolet et al.~\cite{Nicolet2021Large} on surface quality on \textbf{Amphora (Genus-2)}. Less noisy curvature and more continuous reflection lines indicate higher surface quality.}

    \label{fig:surface_detail}

\end{figure}

%% file: sec/6_conclusion.tex
\section{Conclusion}

We demonstrate that our topology-informed inverse rendering method can accurately reconstruct mesh surfaces from multi-view images, capturing essential topological features for high-genus surfaces while enhancing surface details for low-genus surfaces. Both qualitative and quantitative results show that our approach outperforms existing methods in achieving high topological accuracy and detail preservation, with superior performance in Chamfer Distance and Volume IoU, especially for high genus surfaces. We believe our method can provide valuable insights for generative tasks where topological consistency is crucial. Future work will focus on addressing complex topologies with intricate and narrow features, where illumination poses significant challenges, to further advance topology-aware inverse rendering in large-scale generative tasks.